\documentclass[usenatbib]{mn2e}

\usepackage{amsmath}
\usepackage{amssymb}
\usepackage{graphicx}
\usepackage{aas_macros}

\title[CR transport and early impact around SN remnants]{Anisotropic transport and early dynamical impact of Cosmic Rays around Supernova remnants}
\author[Girichidis et al.]{\Large{Philipp~Girichidis$^{1}$\thanks{email: \texttt{philipp@girichidis.com}}}, Thorsten Naab$^{1}$, Stefanie Walch$^{2}$, Micha{\l} Hanasz$^{3}$\\
\vspace*{0.2cm}\\
\scriptsize
$^1$Max Planck Institut f\"{u}r Astrophysik, Karl-Schwarzschild-Str. 1, 85741 Garching, Germany\\
\scriptsize
$^2$Physikalisches Institut I, Universit\"{a}t zu K\"{o}ln, Z\"{u}lpicher Stra{\ss}e 77, 50937 K\"{o}ln, Germany\\
\scriptsize
$^3$Centre for Astronomy, Nicolaus Copernicus University, Faculty of Physics, Astronomy and Informatics, Grudziadzka 5, PL-87100 Toru\'n, Poland}

\newcommand{\rkl}[1]{\left(#1\right)}
\newcommand{\ekl}[1]{\left[#1\right]}

\newcommand{\skl}[1]{\left\langle#1\right\rangle}

\newcommand{\Gauss}{\mathrm{G}}

\newcommand{\Kelvin}{\mathrm{K}}

\newcommand{\pc}{\mathrm{pc}}
\newcommand{\kyr}{\mathrm{kyr}}

\newcommand{\Myr}{\mathrm{Myr}}

\newcommand{\percc}{\mathrm{cm}^{-3}}

\newcommand{\kmpersec}{\mathrm{km}\,\mathrm{s}^{-1}}
\newcommand{\AU}{\mathrm{AU}}
\newcommand{\erg}{\mathrm{erg}}
\newcommand{\cm}{\mathrm{cm}}
\newcommand{\second}{\mathrm{s}}

\newcommand{\diffunit}{\mathrm{cm}^2\,\mathrm{s}^{-1}}
\newcommand{\GeV}{\mathrm{GeV}}

\newcommand{\eV}{\mathrm{eV}}

\newcommand{\delt}{\partial_t}
\newcommand{\delx}{\partial_x}
\newcommand{\encr}{e_{_\mathrm{CR}}}
\newcommand{\encrI}{e_{_{\mathrm{CR},i}}}
\newcommand{\prcr}{p_{_\mathrm{CR}}}
\newcommand{\prcrI}{p_{_{\mathrm{CR},i}}}
\newcommand{\vecv}{\mathbf{v}}
\newcommand{\vecg}{\mathbf{g}}
\newcommand{\vecB}{\mathbf{B}}
\newcommand{\vecF}{\mathbf{F}}
\newcommand{\tenKI}{\mathsf{K}_i}
\newcommand{\QcrI}{Q_{_{\mathrm{CR},i}}}
\newcommand{\tenK}{\mathsf{K}}
\newcommand{\Qcr}{Q_{_\mathrm{CR}}}
\newcommand{\gammaCR}{\gamma_{_\mathrm{CR}}}

\begin{document}

\maketitle
\begin{abstract}
We present a novel implementation of cosmic rays (CR) in the magneto-hydrodynamic code FLASH. CRs are described as separate fluids with different energies. CR advection, energy dependent anisotropic diffusion with respect to the magnetic field and adiabatic losses to follow the evolution of spectra are taken into account. We present a first study of the transport and immediate ($\sim 150\,\kyr$) dynamical impact of CRs on the turbulent magnetised interstellar medium around supernova remnants on scales up to $80\,\pc$. CR diffusion quickly leads to an efficient acceleration of low-density gas (mainly perpendicular to the magnetic field), with accelerations, up to two orders of magnitude above the thermal values. Peaked (at $1\,\GeV$) CR injection spectra have a stronger impact on the dynamics than power-law spectra. For self-consistent magnetic field configurations low energy CRs (with smaller diffusion coefficients) distribute anisotropically with large spatial variations of a factor of ten and more. Adiabatic losses can change the local spectra perceptibly but do not have an integral effect on the dynamics at the spatial and temporal scales considered here. We discuss the potential global impact of CRs and anisotropic transport on the dynamical structure of the ISM and also detail the limitations of the model.
\end{abstract}


\section{Introduction}%
\label{sec:introduction}

Cosmic rays (CR) are high energy particles and constitute an important ingredient of the interstellar medium (ISM) and the Galaxy. Observations find that CR energy densities are comparable to magnetic energy densities and account for a significant overall energy density in the ISM besides thermal and turbulent energy \citep{BoularesCox1990,BeckKrause2005}. Whether CRs can actively regulate processes in the ISM or drive winds and outflows strongly depends on upon their spatial distribution as well as their coupling between with gas. Non-linear processes and scattering effects prevent CRs from streaming freely through the ISM \citep[see reviews by][]{BreitschwerdtEtAl2002, DorfiBreitschwerdt2012, Ferriere2001, Zweibel2013} but they are dynamically coupled to the ISM and can be considered as a fluid under certain assumptions (see below). A key aspect determining the importance of CRs to the dynamics of the ISM is the propagation of CRs from their sites of acceleration through the ISM and the resulting pressure gradients in CR energy density that can contribute to the overall balance of forces in the ISM.

CR are mostly protons and electrons with an observed ratio of about 10:1 proton to electron \citep{LackiThompsonQuataert2010}, however with significant uncertainty. CR energy spectra have been measured over many orders orders of magnitude from $E_\mathrm{CR}\sim10^{7}\,\eV$ up to $\sim10^{20}\,\eV$. In general, the energy spectra peak at around $1\,\GeV$ and are rather steep for higher energies, $N(E)\propto E^{-2.7}$. Therefore, most of the total energy is in CRs with $E_\mathrm{CR}\sim1\,\GeV$, which have to be considered when studying the dynamical impact of CRs on the ISM. Although both electrons and protons are accelerated in strong shocks, protons carry perceptibly more energy than the electrons. In this study we therefore only consider protons when referring to CRs.

The main acceleration mechanism for Galactic CRs is considered to be \emph{diffusive shock acceleration} (DSA) \citep{AxfordEtAl1977, Krymskii1977, Bell1978, BlandfordOstriker1978} and \emph{non-linear DSA} \citep{MalkovOCDrury2001} in shocks of supernova remnants (SNR) (see \citet{Hillas2005} for a review). The DSA model predicts a power-law spectrum in momentum \citep[for a review see, e.g.][]{Blasi2013}. For relativistic particles this translates into $N(E)\propto E^{-s}$, with the scaling exponent, $s$, depending on the shock properties. For very strong shocks the exponent asymptotes to $2$. In the non-relativistic case strong shocks result in $s=3/2$. Even though CRs are treated as tracer particles in the DSA, the particles influence the shock regions and change the shock properties when considering the non-linear extension. These back-reactions results in spectra that are no longer simple power-laws. More recently a number of groups have investigated the shock properties, the CR acceleration, and the CR escape conditions in more detail with numerical simulations \citep{CaprioliEtAl2009, CaprioliBlasiAmato2009, PtuskinEtAl2010, CaprioliEtAl2010, BellSchureReville2011, BellEtAl2013, MorlinoEtAl2013, CaprioliSpitkovsky2014a, CaprioliSpitkovsky2014b}. \citet{TelezhinskyEtAl2012, TelezhinskyEtAl2012b} follow the escape of CRs from SN remnants and their interaction with dense media.

Observations of SN remnants amend the theoretical connection between CRs and shocks. \citet{BambaKoyamaTimoda2000, BorkowskiEtAl2001, VinkEtAl2006} suggest SN remnants to be efficient CR accelerators. If gamma-ray emission could be confirmed to arise from pion production and decay, this would strongly support the paradigm of SNe being the main production sites of CRs in the Galaxy. Recently, \citet{MorlinoEtAl2014} constrain shock dynamics via Balmer line emission from RCW~86 (G315.4-2.3).

Being charged moving particles, CRs are deflected by the magnetic field in the Galaxy. Provided that the field is strong enough to keep the CRs on small gyro-radii relative to Galactic scales, CRs do not stream freely through the ISM. Instead their motions are better described by a diffusion process, where the diffusion along the magnetic field lines is different from the diffusion perpendicular. Determining the diffusive motions theoretically is very complex \citep{Schlickeiser2002,YanLazarian2004,YanLazarian2008,Zweibel2013}. Therefore, observations of CRs are needed to find limits for the diffusion coefficients of CRs. Theoretical considerations suggest a highly anisotropic diffusion with respect to the orientation of the magnetic field. In addition the diffusion depends on the particle energy. For Galactic environments the diffusion coefficient along the magnetic field lines is assumed to be
\begin{equation}
  K_\parallel (E) = K_{\parallel, 0}\,\rkl{\frac{E}{10\,\GeV}}^s,
\end{equation}
where $K_{\parallel, 0}\approx10^{28}-10^{29}\,\cm\,\second^{-1}$ and $s\approx0.3-0.7$ \citep{Berezinskii1990, CastellinaDonato2011, TrottaEtAl2011}. The diffusion coefficient perpendicular to the field lines is probably one or two orders of magnitude smaller \citep{NavaGabici2013, HanaszEtAl2013}, however with significant uncertainty. On spatial scales of the order of the Galaxy, CRs are expected to diffuse almost isotropically \citep[see, e.g.][]{StrongEtAl2007}. Simulations of entire galaxies, investigating the effects of CRs on galactic winds, include CR diffusion in both manners, anisotropically \citep{YangEtAl2012, HanaszEtAl2013} as well as isotropically \citep{SalemBryan2013, BoothEtAl2013}, however in all cases with one global CR fluid, i.e. without energy dependent diffusion coefficients. As the diffusion coefficients and thus the speeds at which CRs diffuse through the ISM are relatively large compared to the turbulent velocities, sound speeds and magnetic waves, it is crucial to include energy dependent diffusion effects when investigating the impact of CRs on the dynamics of the ISM.

In this study we examine the impact of CRs accelerated by a SN remnant on the surrounding ISM. We perform magneto-hydrodynamic simulations in which CRs are incorporated into the hydrodynamic equations as ten separate fluids, representing ten different energy bins ranging from $10^{-2}\,\GeV$ to $10^3\,\GeV$. Anisotropic diffusion of CRs is included with energy dependent diffusion coefficients. Adiabatic losses are included in the model as well, however they only have little impact on the spatial scales ($\sim10^2\,\pc$) and the short time scales ($\sim10^2\,\kyr$) we are looking at. We investigate how CR diffuse through the ISM and how and where they actively accelerate the gas.

\section{Numerical Methods}%
\label{sec:methods-initial-conditions}

\subsection{Standard MHD}%

We solve the combined problem of gas physics and CRs by including the CRs as an additional fluid into the ideal magneto-hydrodynamic (MHD) equations (assuming flux freezing).
\begin{align}
  \label{eq:standard-MHD}
  \frac{\partial\rho}{\partial t} + \nabla\cdot\rkl{\rho\vecv} &= 0\\
  \frac{\partial\rho\vecv}{\partial t} + \nabla\cdot\rkl{\rho\vecv\vecv - \frac{\vecB\vecB}{4\pi}} + \nabla p_\mathrm{gas} &= \rho\vecg\label{eq:mom-HD}\\
  \frac{\partial e_\mathrm{gas}}{\partial t} + \nabla\cdot\ekl{\rkl{e_\mathrm{gas} + p_\mathrm{gas}}\vecv - \frac{\vecB(\vecB\cdot\vecv)}{4\pi}} &= \rho\vecv\cdot\vecg\label{eq:ener-HD}\\
  \frac{\partial\vecB}{\partial t} - \nabla \times \rkl{\vecv\times\vecB} &= 0.
\end{align}
Here, $\rho$ is the gas density, $\vecv$ the velocity, $\vecB$ the magnetic field, $p_\mathrm{gas}$ the gas pressure, $e_\mathrm{gas}$ the energy of the gas, and $\vecg$ is the gravitational acceleration, satisfying the Poisson equation $\Delta\Phi = 4\pi G\rho$, with $\Phi$ being the gravitational potential and $G$ Newton's constant.

\subsection{CR in advection-diffusion approximation}%

\begin{figure}
  \centering
  \includegraphics[width=8cm]{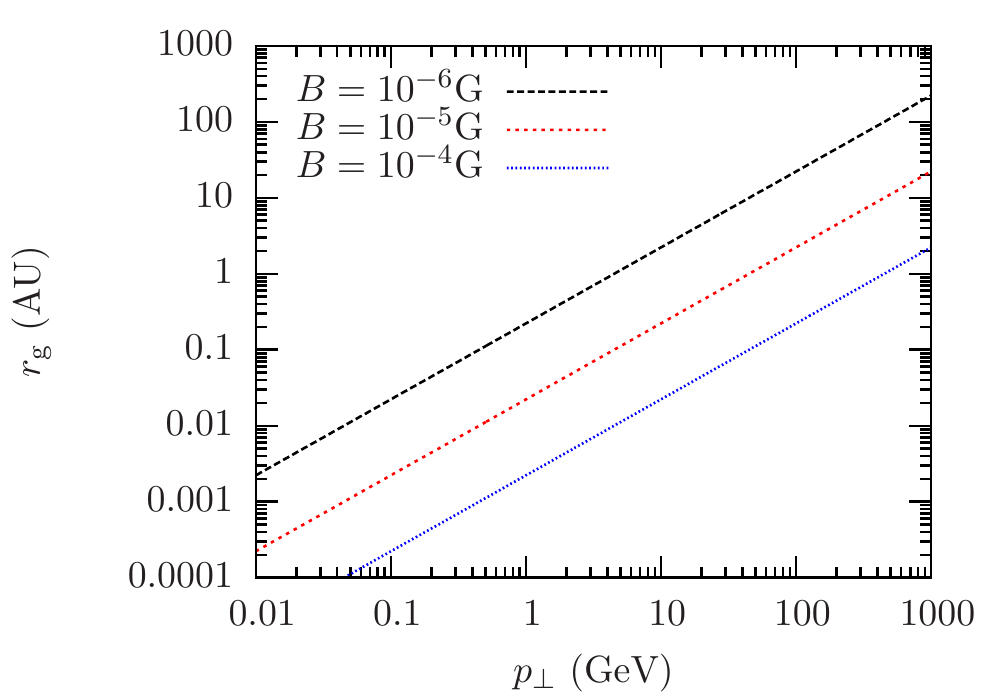}
  \caption{Gyro-radius for CR protons for different magnetic field strengths as a function of CR momentum. The largest gyro-radii (high energy particles in weak magnetic fields) are still small compared to the grid cell resolution of $\Delta x \sim 2\times10^4\,\AU$, which justifies the assumption of a CR fluid.}
  \label{fig:gyroradius}
\end{figure}

Freely moving CRs as relativistic particles can not simply be added to the gaseous system in the hydrodynamic approximation. In order to fulfil the hydrodynamic limit, the particles need to have a mean free path significantly smaller than the computational size of a cell. For a non-relativistic gas, this assumption is easily satisfied. CRs  require a relatively strong magnetic field that couples them to the field lines by redirecting them according to the Lorentz force. The necessary criterion is that the gyro-radius of the CRs $r_\mathrm{gyro,cr} = p_\perp/(|q|B)$ is smaller than the size of the cell in the grid. Here, $p_\perp$ is the momentum of the particles (protons in our case) perpendicular to the magnetic field lines, $q$ is the electric charge, and $B$ is the modulus of the magnetic field. For marginally relativistic particles with a momentum, $p_\perp\sim10^3\,\GeV$, and a weak magnetic field of $B=1\,\mu\Gauss$ the gyro-radius is $r_\mathrm{g}=200\,\AU$ (see Fig.~\ref{fig:gyroradius}). With grid resolutions of the order of $\Delta x \sim 10^{-1}\,\pc\approx2\times10^4\,\AU$, we are in the limit where the hydrodynamical condition is fulfilled.

The CRs which are coupled to the gas exert an additional pressure
\begin{equation}
  \prcr = (\gammaCR-1)\encr,
\end{equation}
where $\gammaCR$ is the adiabatic index for the CR fluid and $\encr$ is the CR energy. As we are investigating a large range of energies, we vary $\gammaCR$ from $\gammaCR=1.6$ for the low energies ($E_\mathrm{CR}=10^{-2}\,\GeV$) up to $\gammaCR=4/3$ for the high energy CRs ($E_\mathrm{CR}=10^{3}\,\GeV$). The CRs need to be added to equations~(\ref{eq:mom-HD}) and (\ref{eq:ener-HD}), where the CR and gas energy (pressure) add up to a total energy (pressure). In addition, the CRs can diffuse (anisotropically with different diffusion coefficients parallel and perpendicular to the magnetic field lines), which means that the energy equation needs to be modified by a diffusion process. We solve the evolution of the energy separately for the CRs as well as for the total energy (gas plus CRs).

We describe the transport of CRs in the ISM with a diffusion-advection approximation following \citet{SchlickeiserLerche1985}
\begin{equation}
  \label{eq:CR-adv-diff-glob}
  \delt \encr + \nabla\cdot(\encr\vecv) = -\prcr\nabla\cdot\vecv + \nabla\cdot(\tenK\nabla\encr) + \Qcr.
\end{equation}
Here, $\encr$ is the CR energy density, $\vecv$ is the gas velocity, $\tenK$ is the CR diffusion tensor, and $\Qcr$ is a CR source term, representing CR energy input (see section~\ref{sec:CR-implementation} for details of the numerical implementation and section~\ref{sec:CR-tests} for tests of the individual components). The diffusion is treated in an anisotropic way \citep[see][]{RyuEtAl2003}
\begin{equation}
  \label{eq:diff-tensor}
  \tenK \equiv K_{ij} = K_\perp \delta_{ij} + (K_\parallel - K_\perp) n_i n_j, \qquad n_i = \frac{B_i}{|\mathbf{B}|},
\end{equation}
with $K_\perp$ and $K_\parallel$ being the diffusion coefficients perpendicular and parallel to the direction of the magnetic field, motivated by numerical experiments by \cite{Jokipii1999, GiacaloneJokipii1999}.

\subsection{Extension to many CR energy bins}%

The general approach described above combines all the CR energy in one energy bin. However, the CR diffusion coefficient depends on the energy, $\tenK\propto\encr^{-0.3}$ \citep[consistent with e.g.][]{NavaGabici2013}. Therefore, we split the total energy equation~(\ref{eq:CR-adv-diff-glob}) into separate energy regimes, $i$,
\begin{equation}
  \label{eq:CR-adv-diff-ebins}
  \delt \encrI + \nabla\cdot(\encrI\vecv) = -\prcrI\nabla\cdot\vecv + \nabla\cdot(\tenKI(\encrI)\nabla\encrI) + \QcrI,
\end{equation}
with different diffusion coefficients $\tenK(\encrI)$, where $\tenK$ has the functional form given in equation~(\ref{eq:diff-tensor}). The perpendicular component of the diffusion tensor is assumed to be two orders of magnitude smaller than the parallel diffusion coefficients, $K_\perp = 0.01\,K_\parallel$. In our setup we use ten bins to split the CR energy. The total CR energy and pressure are then
\begin{align}
  \encr &= \sum_1^{10}\encrI\\
  \prcr &= \sum_i^{10}\prcrI.
\end{align}

\begin{table}
    \caption{CR parameters}
    \label{tab:CR-coefficients}
    \begin{tabular}{rccc}
      bin & energy $(\GeV)$ & $K_\parallel$ $(\diffunit)$ & $K_\perp$ $(\diffunit)$\\
      \hline
      1  &  $1.0\times10^{-2}$ & $3.2\times10^{26}$ & $3.2\times10^{24}$\\
      2  &  $3.6\times10^{-2}$ & $6.0\times10^{26}$ & $6.0\times10^{24}$\\
      3  &  $1.3\times10^{-1}$ & $1.1\times10^{27}$ & $1.1\times10^{25}$\\
      4  &  $4.6\times10^{-1}$ & $2.2\times10^{27}$ & $2.2\times10^{25}$\\
      5  &  $1.7\times10^{+0}$ & $4.1\times10^{27}$ & $4.1\times10^{25}$\\
      6  &  $6.0\times10^{+0}$ & $7.7\times10^{27}$ & $7.7\times10^{25}$\\
      7  &  $2.2\times10^{+1}$ & $1.5\times10^{28}$ & $1.5\times10^{26}$\\
      8  &  $7.7\times10^{+1}$ & $2.8\times10^{28}$ & $2.8\times10^{26}$\\
      9  &  $2.8\times10^{+2}$ & $5.3\times10^{28}$ & $5.3\times10^{26}$\\
      10 &  $1.0\times10^{+3}$ & $1.0\times10^{29}$ & $1.0\times10^{27}$\\
    \end{tabular}

    \medskip
    Cosmic ray energy bins and diffusion coefficients.
\end{table}

We set the energy bins in the range of $\encr\in(10^{-2}-10^{3})\,\GeV$ with the largest parallel diffusion coefficient being $K_\parallel=10^{29}\,\diffunit$. An overview of all CR parameters is shown in Tab.~\ref{tab:CR-coefficients}.

\subsection{Adiabatic losses}%

Including several energy bins for the CRs requires to take into account the shifts in energy space due to compression or expansion of the fluid, i.e. adiabatic gains and losses. Adiabatic losses are included in the energy equations~(\ref{eq:CR-adv-diff-ebins}) for each energy bin. However, compression or expansion of the combined fluid changes the energy of the CRs, which will result in energy shifts between the bins. This needs to be computed separately. If the spectrum is a perfect power-law, adiabatic losses do not change the spectral shape. Contrary, the shape of more complicated and dynamically evolving spectra is changed by adiabatic losses. We treat adiabatic losses similar to \citet{JonesEtAl1999, Miniati2001}. We define the CR number density, $n_i$, and the values of the particle distribution function, $f_i$, as cell centered quantities in energy space ($i=1..10$ for the 10 CR energy bins). The values for the momentum per CR proton, $q_{i-1/2}$, the corresponding energy per CR, $E_{i-1/2}$, as well as the energy fluxes between the bins, $\Phi_{i-1/2}$, are defined at cell boundaries. We emphasise that in this section all momenta are labelled with the letter $q$ in order to distinguish them from the pressure.
Assuming a piecewise constant particle distribution function
\begin{equation}
  f(q) = f_i~,q\in\left[q_{i-1/2},q_{i+1/2}\right)
\end{equation}
gives the particle number density,
\begin{align}
  \label{eq:CR-particle-number}
  n_i &= \int_{q_{i-1/2}}^{q_{i+1/2}} 4\pi q^2 f(q) dq\\
  &= \frac{4\pi}{3}f_i\rkl{q_{i+1/2}^3-q_{i-1/2}^3},
\end{align}
and the particle energy density
\begin{align}
  e_i &= \int_{q_{i-1/2}}^{q_{i+1/2}} E(q)\,4\pi q^2 f(q) dq.
\end{align}
with $E(q)$ being the energy per CR particle.
The adiabatic losses are given by
\begin{equation}
  b(E) \equiv -\rkl{\frac{dE}{dt}}_\mathrm{ad} = (\gammaCR-1)(\nabla\cdot\mathbf{v})E,
\end{equation}
with $\gammaCR-1=0.6$ in the non-relativistic and $\gammaCR-1=1/3$ in the ultra-relativistic regime, which we change linearly with the log of the energy. The changes in energy density over one time step can be written as
\begin{equation}
  e_i^{t+\Delta t} = e_i^t - \Delta t\rkl{\Phi_{i-1/2}-\Phi_{i+1/2}},
\end{equation}
where energy flux $\Phi_{i-1/2}$ within one hydrodynamical time step is given by 
\begin{equation}
  \Phi_{i-1/2} = \frac{1}{\Delta t} \int_t^{t+\Delta t}\,b(E)\,E\, 4\pi q^2\, \left.f(t,q)\right|_{q_{i-1/2}}\,dt.
\end{equation}
This flux can be written as
\begin{equation}
  \label{eq:energy-flux}
  \Phi_{i-1/2} = \frac{4\pi}{\Delta t} \int_{q_{i-1/2}}^{q_u} \,E\, q^2 f_j(q)\,dq,
\end{equation}
where
\begin{equation}
  j = 
  \begin{cases}
    i+1 & \mathrm{if}~b(E) > 0,\\
    i   & \mathrm{if}~b(E) \le 0,
  \end{cases}
\end{equation}
and $q_\mathrm{u}$ is the upstream momentum. The corresponding upstream energy, $E_\mathrm{u}$, respectively, is solution of the integral
\begin{align}
  \Delta t &= - \int_{E_{i-1/2}}^{E_u} \frac{dE}{b(E)}\label{eq:delta-time-energy}\\
  &= - \int_{E_{i-1/2}}^{E_u} \ekl{(\gammaCR-1)\rkl{\nabla\cdot\mathbf{v}}E} ^{-1}\, dE\\
  &= - \frac{1}{(\gammaCR-1)\rkl{\nabla\cdot\mathbf{v}}} \,\ln\rkl{\frac{E_u}{E_{i-1/2}}}.
\end{align}
We can now solve for $E_u$,
\begin{equation}
  E_u = E_{i-1/2}\,\exp\rkl{-(\gammaCR-1)\rkl{\nabla\cdot\mathbf{v}}\,\Delta t},
\end{equation}
and compute Eq.~(\ref{eq:energy-flux}). At this point we need the relation between the kinetic energy of the CRs and their momentum. The CR energies considered here range from the classical limit to the relativistic limit, so the general relation between the kinetic energy and the momentum has to be used,
\begin{align}
  qc &= \sqrt{E^2+2EE_0},
\end{align}
with $E_0$ being the rest energy of a proton. The integral then transforms to
\begin{align}
  \Phi_{i-1/2} &= \frac{4\pi}{c^3\Delta t}\,f_j\,\int_{E_{i-1/2}}^{E_u}  E(E+E_0)(E^2+2EE_0)^{1/2}\, dE
\end{align}
with a lengthy but simple analytical solution.

The equations above can only be computed if we know the number density of the CRs or the distribution function. The number density of CRs can be computed using the CR energy density in the cell, $e_i$, and the average energy per CR, $\skl{E}_i$,
\begin{equation}
  n_i = \frac{E_{i}}{\skl{E}_i}.
\end{equation}
The average energy per CR particle in bin $i$ does not depend on the particle distribution function, as long as the distribution function is piecewise constant,
\begin{equation}
  \label{eq:average-CR-energy}
  \skl{E_\mathrm{CR}}_i = \frac{e_{i}}{n_{i}} = \frac{\int_{q_{i-1/2}}^{q_{i+1/2}}f_id\mathbf{q}}{\int_{q_{i-1/2}}^{q_{i+1/2}}Ef_id\mathbf{q}} = \frac{\int_{q_{i-1/2}}^{q_{i+1/2}}d\mathbf{q}}{\int_{q_{i-1/2}}^{q_{i+1/2}}Ed\mathbf{q}},
\end{equation}
with the total energy being the integrated energy in bin $i$ and the total number being the integrated number in bin $i$, respectively. Knowing $n_i$, we can compute $f_i$ using equation~(\ref{eq:CR-particle-number}) and compute the energy fluxes.

For the integration in energy space a numerical stability criterion similar to the one used for spatial integration applies. We have to ensure that within one time step, $\Delta t$, no CRs are transported for more than one energy bin, i.e. $|E_u-E_{i-1/2}|$ must be less than the energy bin width. We can invert the problem using equation~(\ref{eq:delta-time-energy}): the maximum time step that the energy integration can have is $\Delta t(E_u = E_{i+1/2})$ or $\Delta t(E_u = E_{i-1/2})$ for converging or diverging flows in the local cell, respectively. If $\Delta t < \Delta t_\mathrm{sim}$ then we use sub-cycling in the energy integration. Numerical tests and a discussion of the limits in this numerical model are discussed in section~\ref{sec:ad-losses-tests}.

\subsection{Hydrodynamics and CR fluid}%

Having discussed the MHD equations and the CR fluid, we now need to combine these two fluids. The continuity equation is not influenced at all because the CRs are not implemented with a separate density and velocity field. The momentum equation now contains thermal, magnetic, and CR pressure contributions
\begin{align}\label{eq:total-pressure}
  p_\mathrm{tot} &= p_\mathrm{th} & + &\,\prcr & + &\,p_\mathrm{mag}\\
  &= (\gamma-1)e_\mathrm{th} & + &\,(\gammaCR-1)\encr & + &\,B^2/8\pi.
\end{align}
The closure relation for the system, the equation of state, combines the different contributions from CR and thermal pressure in an effective adiabatic index, $\gamma_\mathrm{eff}$,
\begin{equation}
  \gamma_\mathrm{eff} = \frac{\gamma p_\mathrm{th} + \gammaCR \prcr}{p_\mathrm{th} + \prcr}.
\end{equation}

The combined system of equations that we solve numerically is then given by
\begin{align}
  \frac{\partial\rho}{\partial t} + \nabla\cdot\rkl{\rho\vecv} &= 0\\
  \frac{\partial\rho\vecv}{\partial t} + \nabla\cdot\rkl{\rho\vecv\vecv - \frac{\vecB\vecB}{4\pi}} + \nabla p_\mathrm{tot} &= \rho\vecg\\
  \frac{\partial e}{\partial t} + \nabla\cdot\ekl{\rkl{e + p_\mathrm{tot}}\vecv - \frac{\vecB(\vecB\cdot\vecv)}{4\pi}} &= \rho\vecv\cdot\vecg + \nabla\cdot\tenK\nabla\encr\\
  \frac{\partial\vecB}{\partial t} - \nabla \times \rkl{\vecv\times\vecB} &= 0\\
  \frac{\partial\encrI}{\partial t} + \nabla\cdot(\encrI\vecv) &= -\prcrI\nabla\cdot\vecv\\
  &\phantom{=}+ \nabla\cdot(\tenKI\nabla\encrI)\notag\\
  &\phantom{=}+ \QcrI \qquad i\in 1..10.\notag,
\end{align}
where the total energy is given by
\begin{equation}
    e = 0.5\rho v^2 + e_\mathrm{th} + \encr + B^2/8\pi.
\end{equation}

\begin{figure}
  \centering
  \includegraphics[width=8cm]{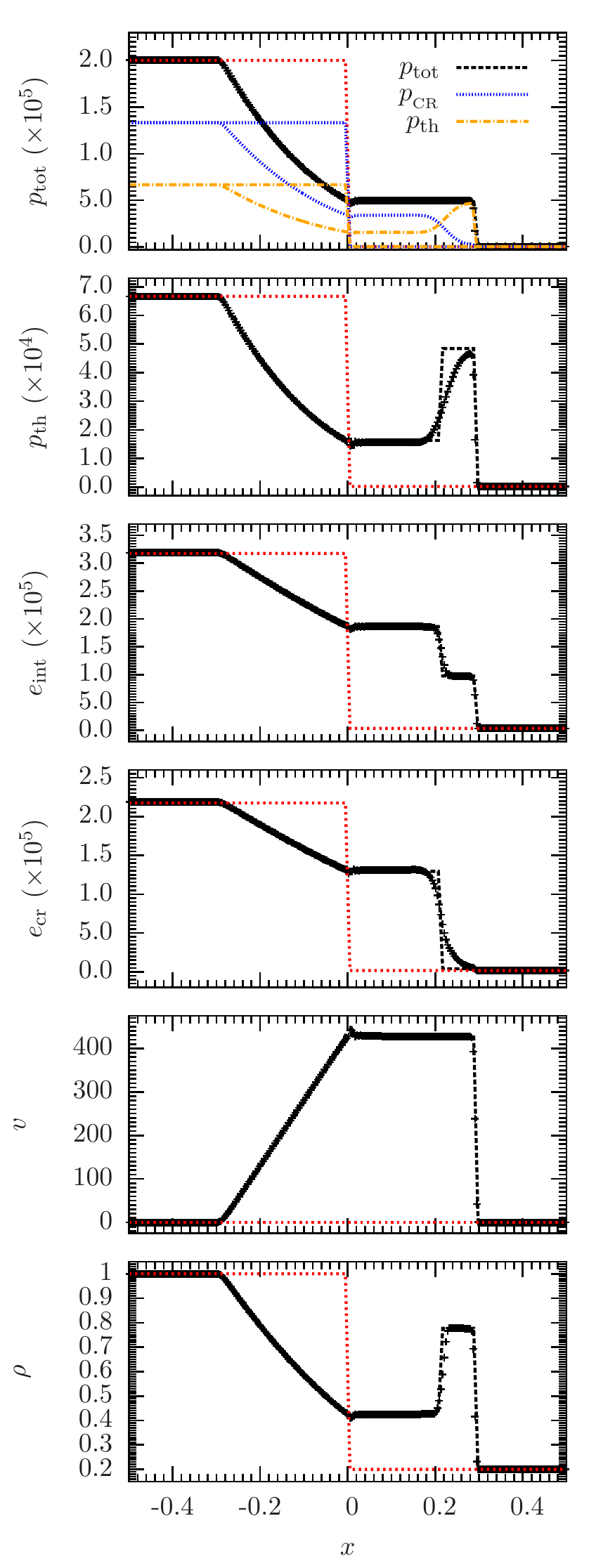}
  \caption{Sod shock tube for the composite fluid of gas and CRs. The black dotted lines show the analytic solution, the crosses the data points from the combined MHD and CR solver. The red line indicates the initial configuration at $t=0$. Although the main features of the shock are captured, the numerical solution shows visual deviations from the analytic solution.}
  \label{fig:Sod-test}
\end{figure}

We implement the solver in the astrophysical code \textsc{Flash} \citep{FLASH00, DubeyEtAl2008} in version 4. The underlying MHD solver is the HLL3R solver \citet{Bouchut2007, Bouchut2010, Waagan2009, Waagan2011}, which we extend with the CR fluids. We perform several tests of the individual components, which are shown in the appendix. The Sod shock tube test \citep{Sod1978} was extended to a two-component fluid including CRs by \citet{PfrommerEtAl2006} and is shown in Fig.~\ref{fig:Sod-test}. The main features of the shock are captured. However, as the solver is quite diffusive, the shock front itself shows noticeable deviations from the analytical solution. Similar deviations are therefore also seen for the solver without CRs.

\subsection{Injection of CRs}%

As presented in the introduction, CRs are accelerated in shocks. Ideally, CRs would be implemented as a source $\QcrI$ depending on the shock properties of the SN remnant. However, the resolution in our setup is too low to accurately determine the shock properties. We therefore define a spherical injection region with a radius of $5\,\pc$, in which we inject the thermal SN energy as well as the CR energy CRs uniformly. Despite its simplicity, our implementation might qualitatively behave in a similar way to the detailed shock acceleration models. \citet{BellEtAl2013} report that the majority of the CRs accelerated in the SN shock do not escape into the upstream region but move downstream into the SN remnant. The confined CRs can thus undergo further acceleration cycles by passing the shock region again.

\section{Numerical setup and initial conditions}%

We investigate the impact of CRs from a SNR on the ISM and follow the evolution of the CR energy spectra over time. To do this, we set up a cubic box with an edge length of $L_\mathrm{box}=80\,\pc$, in whose centre we place a SN explosion by injecting $E_\mathrm{SN}$ of thermal energy within a sphere of radius $5\,\pc$. Given the low resolution of $128^3$ cells, we refrain from choosing a smaller injection radius to avoid numerical grid effects. A fraction of 10\% and 30\% of the SN energy is injected in the form of CRs. Therefore, we compare every run with CRs ($E_\mathrm{CR}=f_\mathrm{CR}E_\mathrm{SN}$, $E_\mathrm{SN}=10^{51}\,\erg$) with a SN energy input of $E_\mathrm{SN}=(1+f_\mathrm{CR})10^{51}\,\erg$ and $E_\mathrm{CR}=0$ (see table~\ref{tab:simulations}). The CR source spectrum is (see references in the introduction)
\begin{equation}
  \label{eq:source-spectrum1}
  N_\mathrm{CR}(E) \propto E^{-2},
\end{equation}
motivated by shock acceleration models. In addition we also perform separate simulations with a source spectrum
\begin{equation}
  \label{eq:source-spectrum2}
  N_\mathrm{CR}(E) \propto
  \begin{cases}
    E^2 & \mathrm{for~} E\le1\,\GeV\\
    E^{-2} & \mathrm{for~} E>1\,\GeV,
  \end{cases}
\end{equation}
for comparison. This spectrum resembles the peak at $E_\mathrm{CR}\approx1\,\GeV$.

\begin{figure*}
  \begin{minipage}{\textwidth}
    \centering
    \includegraphics[width=5cm]{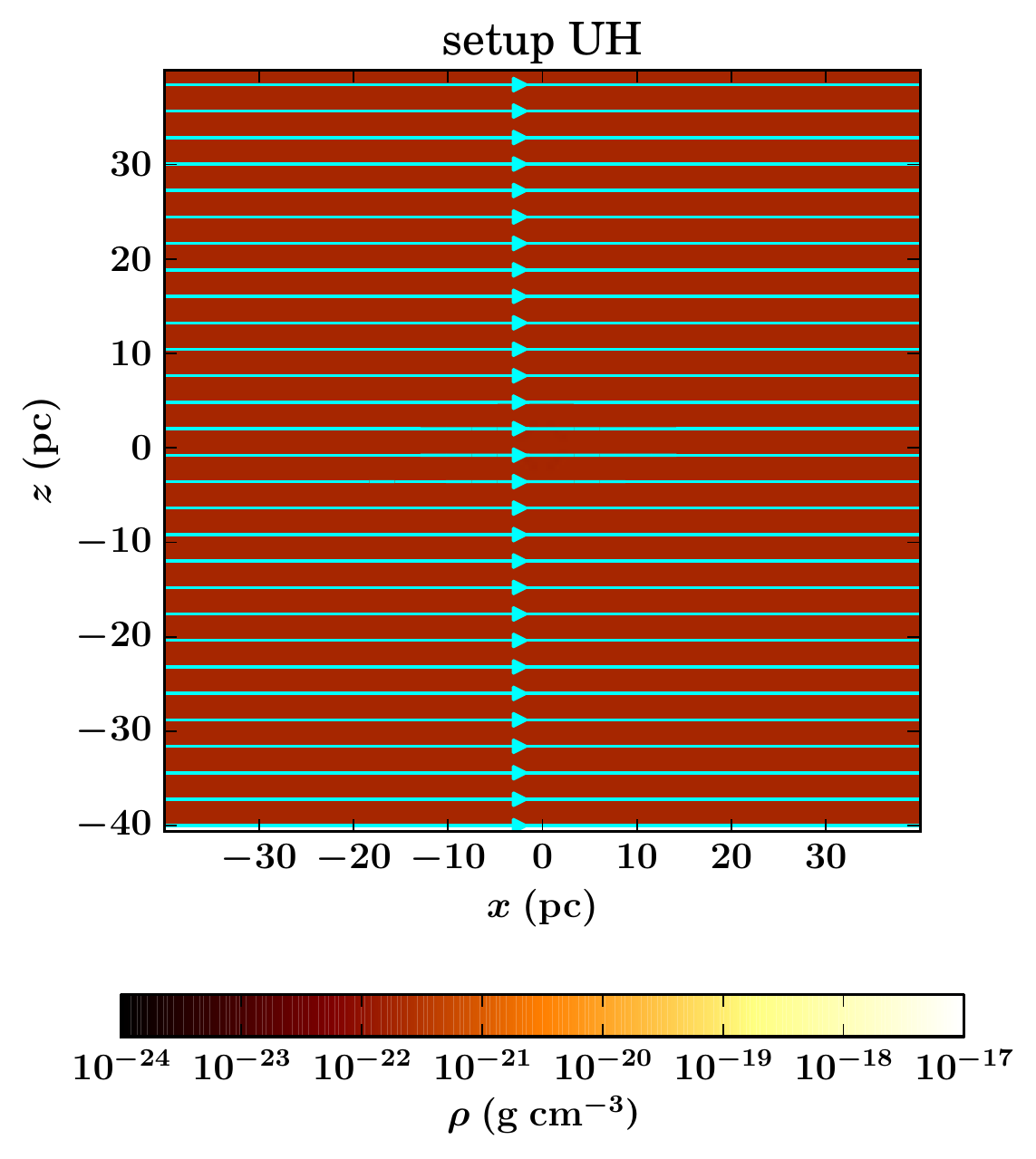}
    \includegraphics[width=5cm]{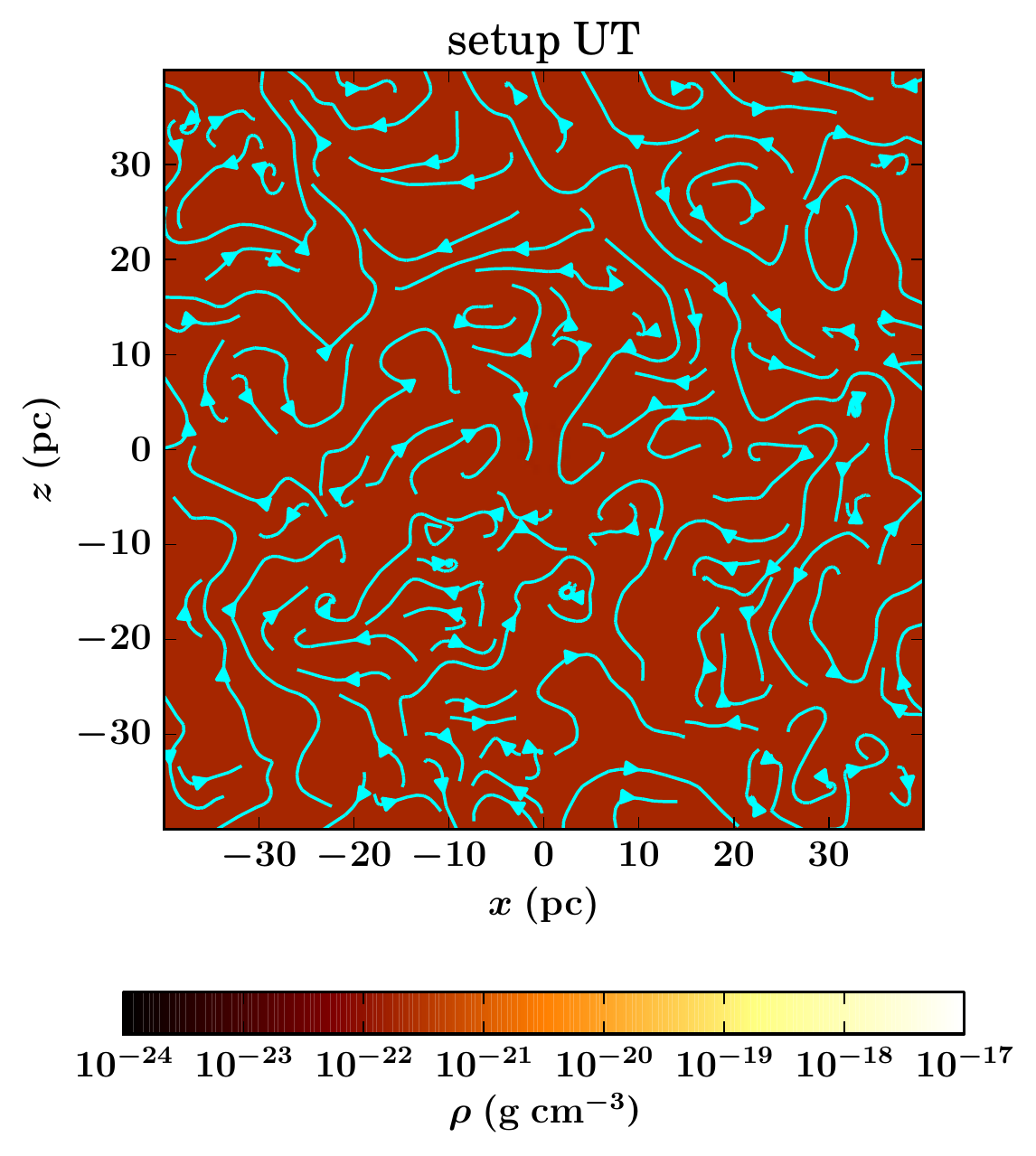}
    \includegraphics[width=5cm]{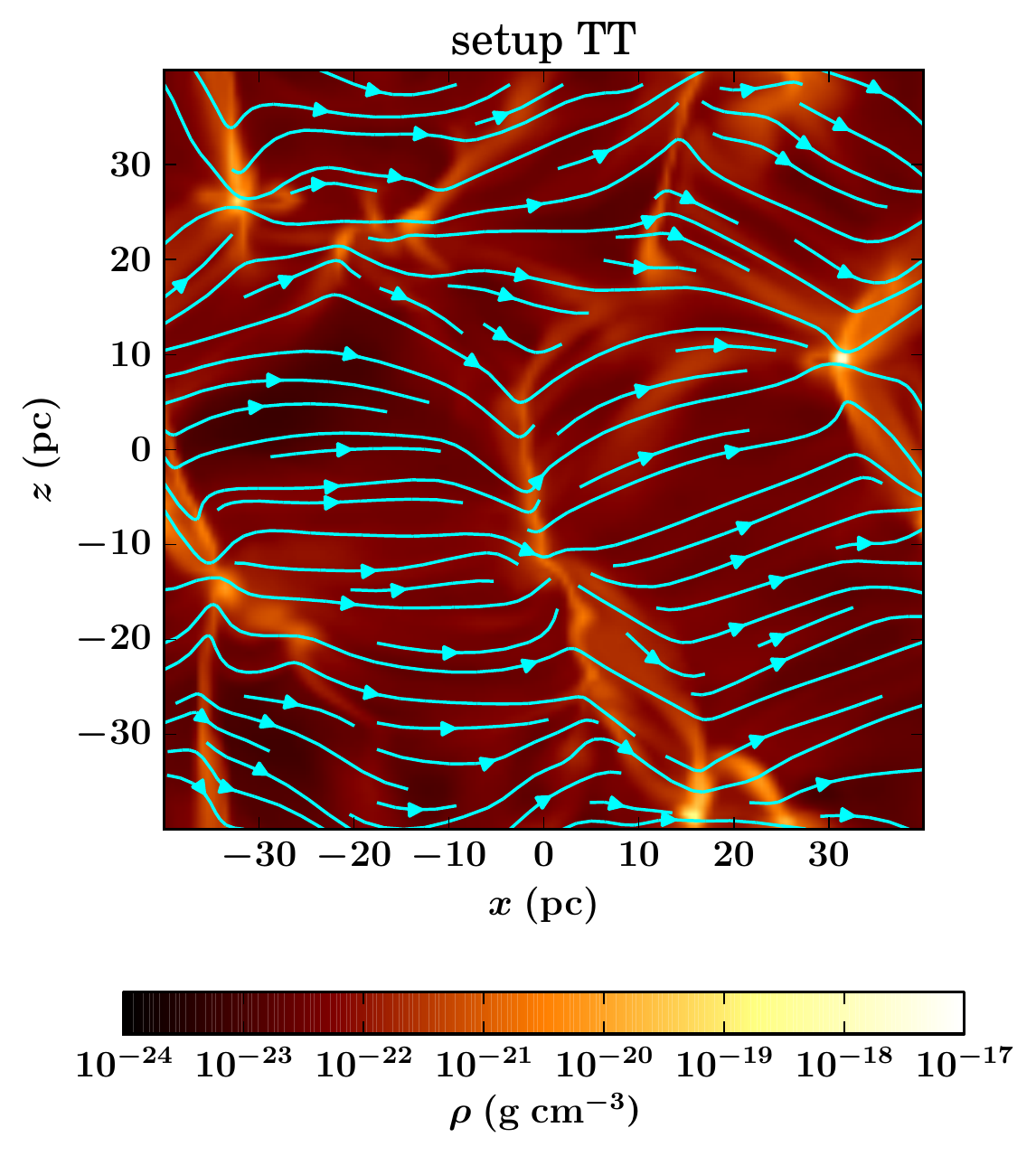}
    \caption{Slices with the density and structure of the magnetic field lines just before the explosion of the SN: a homogeneous density ($\rho=10\,\percc$) with magnetic field lines in $x$ direction (UH, left), a homogeneous density ($\rho=10\,\percc$) with tangled magnetic field (UT, centre), and a self-consistent turbulent ISM (TT, right).}
    \label{fig:init-dens-mag}
  \end{minipage}
\end{figure*}

\begin{table*}
  \caption{Overview of the simulations}
  \label{tab:simulations}
  \begin{minipage}{0.7\textwidth}
    \begin{center}
    \begin{tabular}{lccccccc}
      name & setup & magnetic & density & $f_\mathrm{CR}$ & $E_\mathrm{SN}$ & spectral peak & incl. ad.\\
       &  &  field & & & (erg) & ($\GeV$) & losses\\
      \hline
      \texttt{UH-SN1$\!$.$\!$0-CR0$\!$.$\!$1-PL} & UH & hom.  & unif. & 0.1 & $1.0\times10^{51}$ & $10^{-2}$ & no\\
      \hline                                                              
      \texttt{UT-SN1$\!$.$\!$0-CR0$\!$.$\!$1-PL} & UT & tan.  & unif. & 0.1 & $1.0\times10^{51}$ & $10^{-2}$ & no\\
      \hline                                                              
      \texttt{TT-SN1$\!$.$\!$0-CR0$\!$.$\!$0} & TT & turb. & turb. & 0.0 & $1.0\times10^{51}$ & $-$ & $-$\\
      \texttt{TT-SN1$\!$.$\!$1-CR0$\!$.$\!$0} & TT & turb. & turb. & 0.0 & $1.1\times10^{51}$ & $-$ & $-$\\
      \texttt{TT-SN1$\!$.$\!$3-CR0$\!$.$\!$0} & TT & turb. & turb. & 0.0 & $1.3\times10^{51}$ & $-$ & $-$\\
      \hline
      \texttt{TT-SN1$\!$.$\!$0-CR0$\!$.$\!$1-PL}    & TT & turb. & turb. & 0.1 & $1.0\times10^{51}$ & $10^{-2}$ & no\\
      \texttt{TT-SN1$\!$.$\!$0-CR0$\!$.$\!$3-PL}    & TT & turb. & turb. & 0.3 & $1.0\times10^{51}$ & $10^{-2}$ & no\\
      \texttt{TT-SN1$\!$.$\!$0-CR0$\!$.$\!$1-PL-ad} & TT & turb. & turb. & 0.1 & $1.0\times10^{51}$ & $10^{-2}$ & yes\\
      \texttt{TT-SN1$\!$.$\!$0-CR0$\!$.$\!$3-PL-ad} & TT & turb. & turb. & 0.3 & $1.0\times10^{51}$ & $10^{-2}$ & yes\\
      \hline                                                              
      \texttt{TT-SN1$\!$.$\!$0-CR0$\!$.$\!$1-PK}    & TT & turb. & turb. & 0.1 & $1.0\times10^{51}$ & $1$ & no\\
      \texttt{TT-SN1$\!$.$\!$0-CR0$\!$.$\!$3-PK}    & TT & turb. & turb. & 0.3 & $1.0\times10^{51}$ & $1$ & no\\
      \texttt{TT-SN1$\!$.$\!$0-CR0$\!$.$\!$1-PK-ad} & TT & turb. & turb. & 0.1 & $1.0\times10^{51}$ & $1$ & yes\\
      \texttt{TT-SN1$\!$.$\!$0-CR0$\!$.$\!$3-PK-ad} & TT & turb. & turb. & 0.3 & $1.0\times10^{51}$ & $1$ & yes\\
      \hline
    \end{tabular}
  \end{center}
  \medskip
  
  Listed are the name of the simulation, the setup followed by the magnetic field configuration and the density structure. The factor $f_\mathrm{CR}$ gives the fraction of CR energy, $E_\mathrm{SN}$ the total injected energy (thermal or thermal plus CRs). The last two columns specify the shape of the spectrum and whether adiabatic losses are included.
  \end{minipage}
\end{table*}

We follow the evolution of CRs with three different setups. We start with two idealised setups of uniform density, one with a homogeneous magnetic field (UH), the other with a tangled field configuration (UT). The main analysis of the paper uses a third setup (TT), where we create a turbulent ISM with a self-consistent magnetic field and density distribution. The configuration of the magnetic fields and the density distributions just before the explosion of the central SNe are shown in Fig.~\ref{fig:init-dens-mag} and listed in Tab.~\ref{tab:simulations}.

\subsection{Setup UH: homogeneous box with homogeneous magnetic field}%

In the first setup (UH; Fig.~\ref{fig:init-dens-mag}, left) we set the magnetic field to be homogeneous along the $x$ axis, $\mathbf{B} = B_0\mathbf{x}$, with $B_0=2\,\mu\Gauss$. The density in the box is constant with $\rho=10\,\percc$. The SN explodes at $t=0$. The gas is isothermal at the point of SN injection with a temperature of $T=30\,\Kelvin$ and behaves adiabatically with $\gamma=5/3$.

\subsection{Setup UT: homogeneous box with tangled magnetic field}%

Also setup UT (Fig.~\ref{fig:init-dens-mag}, centre) has a homogeneous density distribution with $\rho=10\,\percc$ and a tangled magnetic field. We generate the magnetic field in Fourier space with an isotropic power spectrum of the form
\begin{equation}
  \mathbf{\tilde{B}}(|\mathbf{k}|) \propto
  \begin{cases}
    k^3 & \mathrm{for~} k\le4\\
    k^{-3} & \mathrm{for~} k>4,
  \end{cases}
\end{equation}
where $k=1$ corresponds to the size of the box. The exact values for the scaling exponents do not have a physical motivation here. We chose a such steep power-laws to strongly populate modes with $k\approx4$. The vectors are projected in $k$ to give a divergence-free field $\mathbf{B}(\mathrm{x})$ in real space. The normalisation of the magnetic field is chosen such that the r.m.s. value of the field yields $2\,\mu\Gauss$. The SN ignites at $t=0$.

\subsection{Setup TT: structured ISM with a self-consistently evolved magnetic field}%

Class TT (Fig.~\ref{fig:init-dens-mag}, right) of the simulation setups uses a structured ISM with self-consistently evolved magnetic fields. We create these conditions by setting up a periodic box with a uniform density and a homogeneous magnetic field. We then impose a turbulent velocity field, which is generated in Fourier space with an isotropic power spectrum
\begin{equation}
  \mathbf{\tilde{v}}(|\mathbf{k}|) \propto
  \begin{cases}
    k^2 & \mathrm{for~} k\le2\\
    k^{-4} & \mathrm{for~} k>2,
  \end{cases}
\end{equation}
where $k=1$ corresponds again to the size of the box. The scaling exponent for $k>2$, $-4$, is the value for compressible Burgers turbulence in one-dimension. The radially integrated power spectrum then scales as $k^{-2}$. On scales larger than half of the box size ($k<2$) we chose a positive value, i.e. less power on the largest scales, to avoid the formation of only one big clump in the box. Unlike the magnetic field in setup UT we do not project the field in Fourier space, yielding a velocity field in real space that consists of a mixture of compressive and solenoidal modes with a statistical average of 2:1 for solenoidal to compressive modes. The normalisation of the field is chosen such that the initial r.m.s. velocity is $1\,\kmpersec$, which increases to about $5\,\kmpersec$ during the evolution due to the impact of self-gravity. We follow this initial turbulent setup for about $10\,\Myr$ until significant overdensities have formed under the impact of the initial turbulent motions and self-gravity before igniting the SN in the centre of the box.

\section{Results}%
\label{sec:results}

\subsection{General remarks}%

\begin{figure*}
  \begin{minipage}{\textwidth}
    \centering
    \includegraphics[width=16cm]{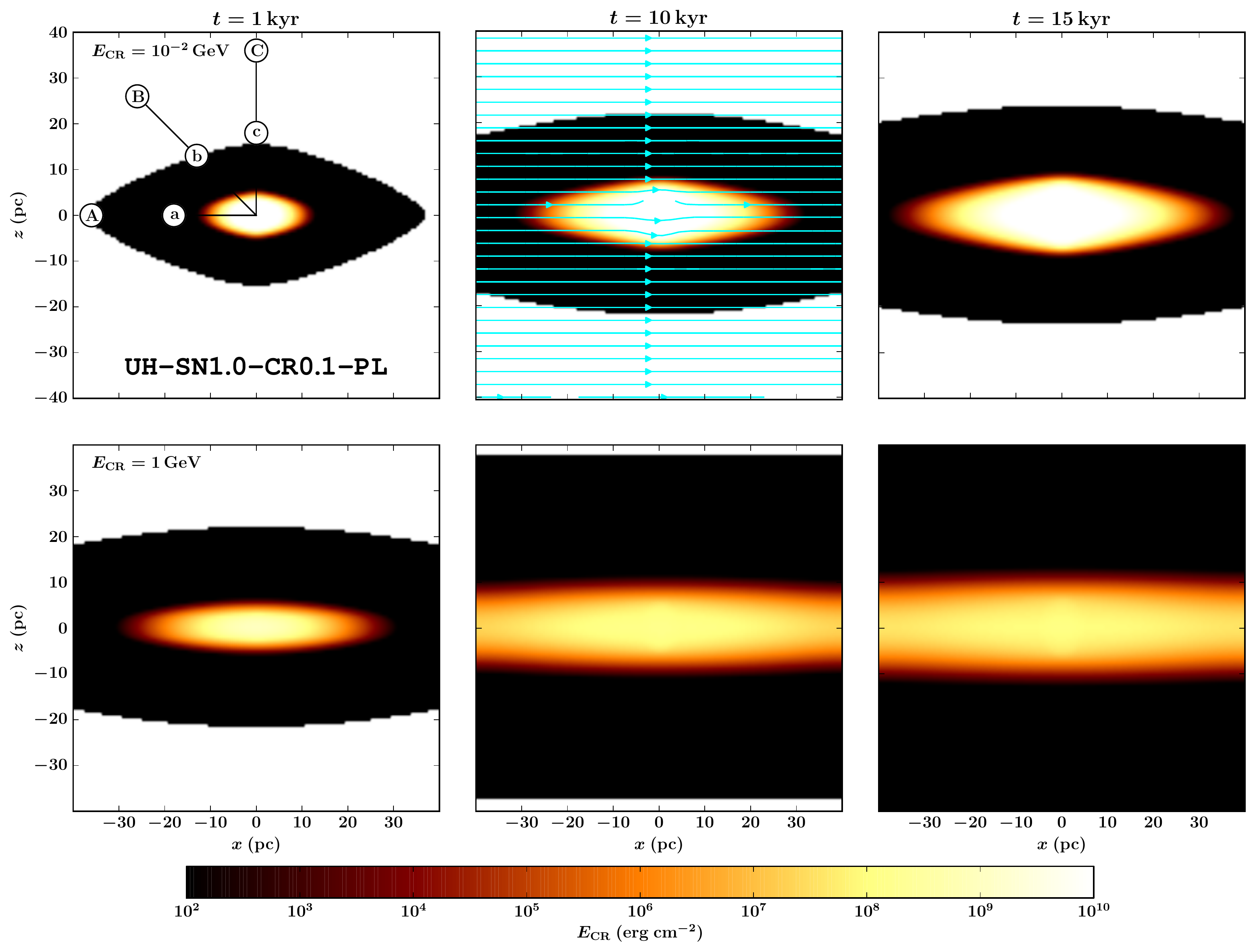}
    \caption{CR diffusion for setup UH with uniform density and homogeneous magnetic field, indicated by the cyan vector field. The columns show different times, $t=1\,\kyr$, $t=10\,\kyr$, and $t=15\,\kyr$. The rows show the projected CR energy for different the energy bins $E_\mathrm{CR}=10^{-2}\,\GeV$ and $E_\mathrm{CR}=1\,\GeV$.
As the source spectrum of the CRs peaks at low energies, the upper panels show more total energy than the lower panels. The diffusion coefficient of the low-energy CRs is roughly one order of magnitude smaller than the coefficient for the high-energy CRs. The strong difference between the coefficients along and perpendicular to the magnetic field lines lead to the highly anisotropic distribution.}
    \label{fig:A-diff-overview}
  \end{minipage}
\end{figure*}

Before we analyse the individual setups in detail, we would like to stress some general properties that are similar for all setups. Due to the large diffusion coefficients the CRs quickly reach the boundaries of the simulated box. At the scales we are investigating in this study, the CR diffusion time scales are shorter than the hydrodynamical time scales. Therefore, we primarily focus on the diffusion properties and do not investigate the long term evolution of the thermal impact of the SN on the gas surrounding it. Nonetheless, the energy transfer from CRs to kinetic energy of the gas is evident even on small spatial ($\sim10^2\,\pc$) and small temporal ($\sim10^2\,\kyr$) scales. High-energy CRs with $10^3\,\GeV$ reach the boundary of the box at only $\sim1\,\kyr$ after the explosion of the SN. Figure~\ref{fig:A-diff-overview} illustrates the CR diffusion for different CR energies and different times ($t=1\,\kyr$, $t=10\,\kyr$, $t=15\,\kyr$) after the explosion of the SN for setup UH. The top row shows the projected CR energy for the lowest energy bin ($E_\mathrm{CR}=10^{-2}\,\GeV$), the bottom panel to bottom row depicts the distribution for intermediate energy CRs ($E_\mathrm{CR}=1\,\GeV$). For CRs above $E_\mathrm{CR}>1\,\GeV$ the distribution of CRs is visually indistinguishable from a cylindrical shape. Due to the large difference between the perpendicular and parallel diffusion coefficients the resulting CR distribution is highly anisotropic. Both the projected CR energy density distribution of the CR energy density as well as the CR spectra show this anisotropy.

\begin{figure}
  \centering
  \includegraphics[width=8cm]{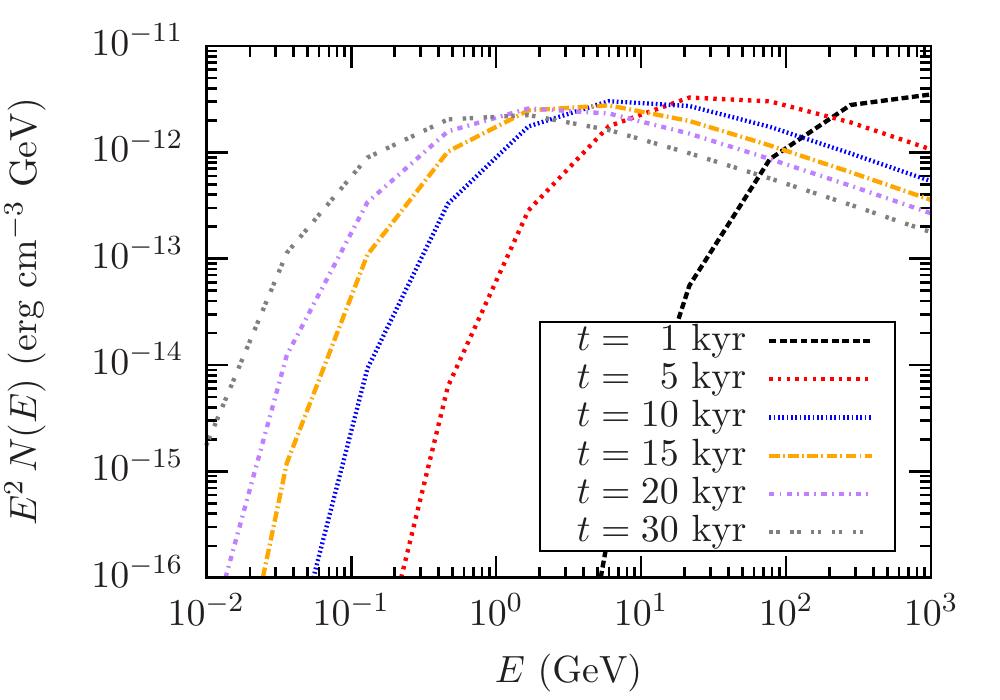}
  \caption{Time evolution of the spectrum for simulation UH at measurement point A. We plot $E^2\,N(E)$, so the source spectrum with $N(E)\propto E^{-2}$ would appear as a horizontal line. The spectrum evolves from the high-energy part due to the larger diffusion coefficients. Over time the high-energy end of the spectrum decreases because a significant amount of the CRs has diffused out of the box. Throughout the simulation the high-energy part of spectrum observed at point A is steeper than the source spectrum.}
  \label{fig:spec-time-evol}
\end{figure}

In this idealised setup we can also see how the spectra evolve over time due to the different diffusion coefficients. Figure~\ref{fig:spec-time-evol} shows how the spectrum at measurement point A (at a distance of $36\,\pc$ from the SN) evolves over time starting with the high-energy part of the CRs. Here we plot $E^2\,N(E)$, so the source spectrum with $N(E)\propto E^{-2}$ would appear as a horizontal line. The peak of the spectrum subsequently shifts to lower energies. At later times, a significant fraction of the high-energy CRs has either left the box or diffused perpendicular to the field lines (along the $y$ and $z$-direction), which leads to a decrease of the total power in the high-energy part of the spectrum. In this simplified magnetic field configuration the high-energy slope of the spectrum at later times when the CRs at peak at energy has passed the measurement point is always steeper than the source spectrum. 

\begin{figure*}
  \begin{minipage}{\textwidth}
    \centering
    \includegraphics[width=16cm]{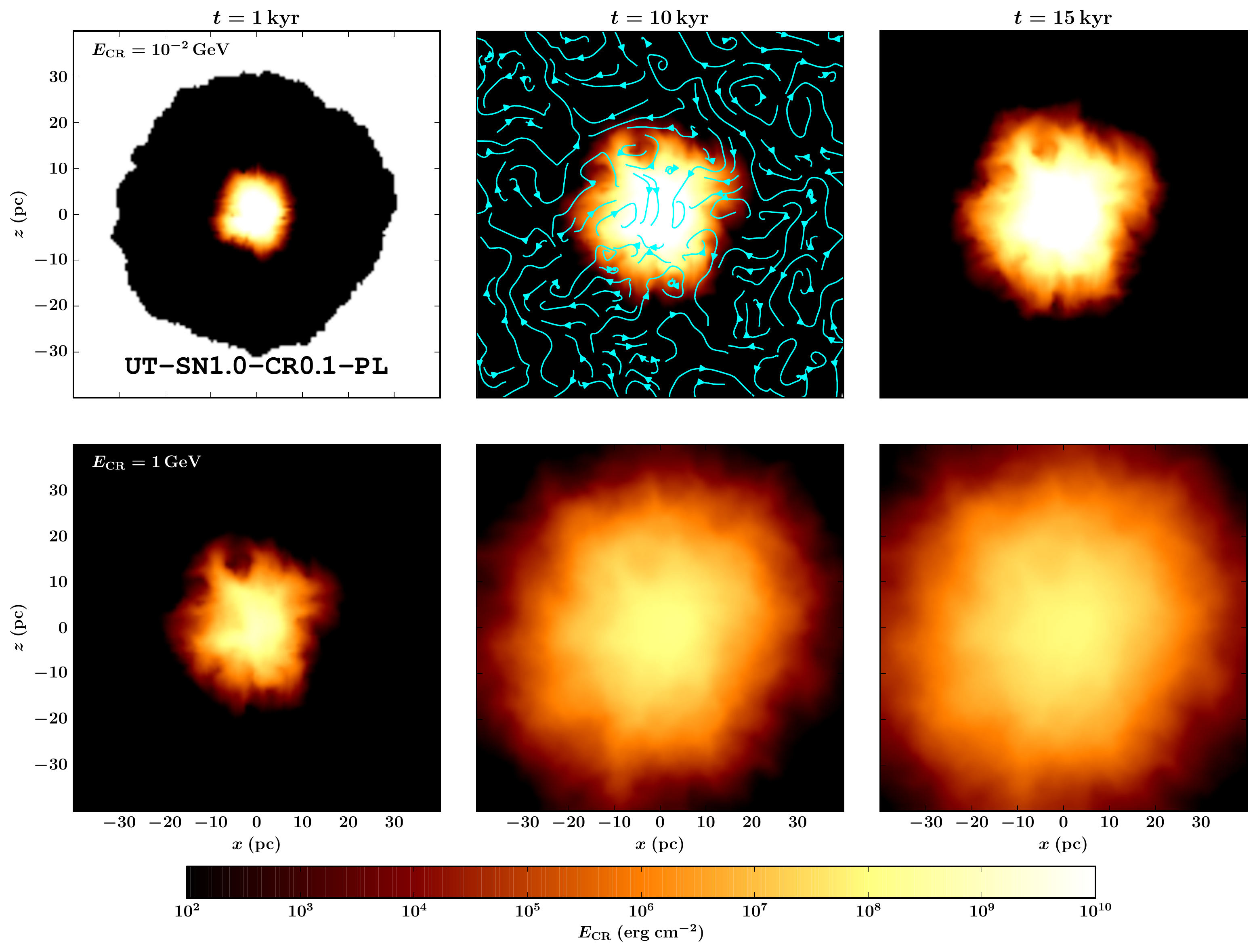}
    \caption{CR diffusion for setup \texttt{UH-SN1$\!$.$\!$0-CR0$\!$.$\!$1-PL} with homogeneous density and tangled magnetic field, indicated by the cyan vector field. The columns represent different times, $t=1\,\kyr$, $t=10\,\kyr$, and $t=15\,\kyr$, the rows show the projected CR energy for different energy bins, $E_\mathrm{CR}=10^{-2}\,\GeV$ and $E_\mathrm{CR}=1\,\GeV$. The tangled structure of the magnetic field leads to an almost spherically symmetric energy distribution and an effective diffusion coefficient between the parallel and perpendicular one.}
    \label{fig:B-diff-overview}
  \end{minipage}
\end{figure*}

The strong differences between the parallel and perpendicular diffusion coefficient are attenuated in the presence of a tangled magnetic field. Fig.~\ref{fig:B-diff-overview} shows the CR energy distribution for setup UT with homogeneous density and tangled magnetic field. The distribution is closer to spherical symmetry since the net CR diffusion is composed of a spatially varying diffusion along and perpendicular to the magnetic field. In this case, the net diffusion could almost be described by an isotropic diffusion process with an effective diffusion coefficient in between the extreme values ($K_\perp<K_\mathrm{eff}<K_\parallel$).

\subsection{Analysis of setups TT}

\begin{figure*}
  \begin{minipage}{\textwidth}
    \centering
    \includegraphics[width=16cm]{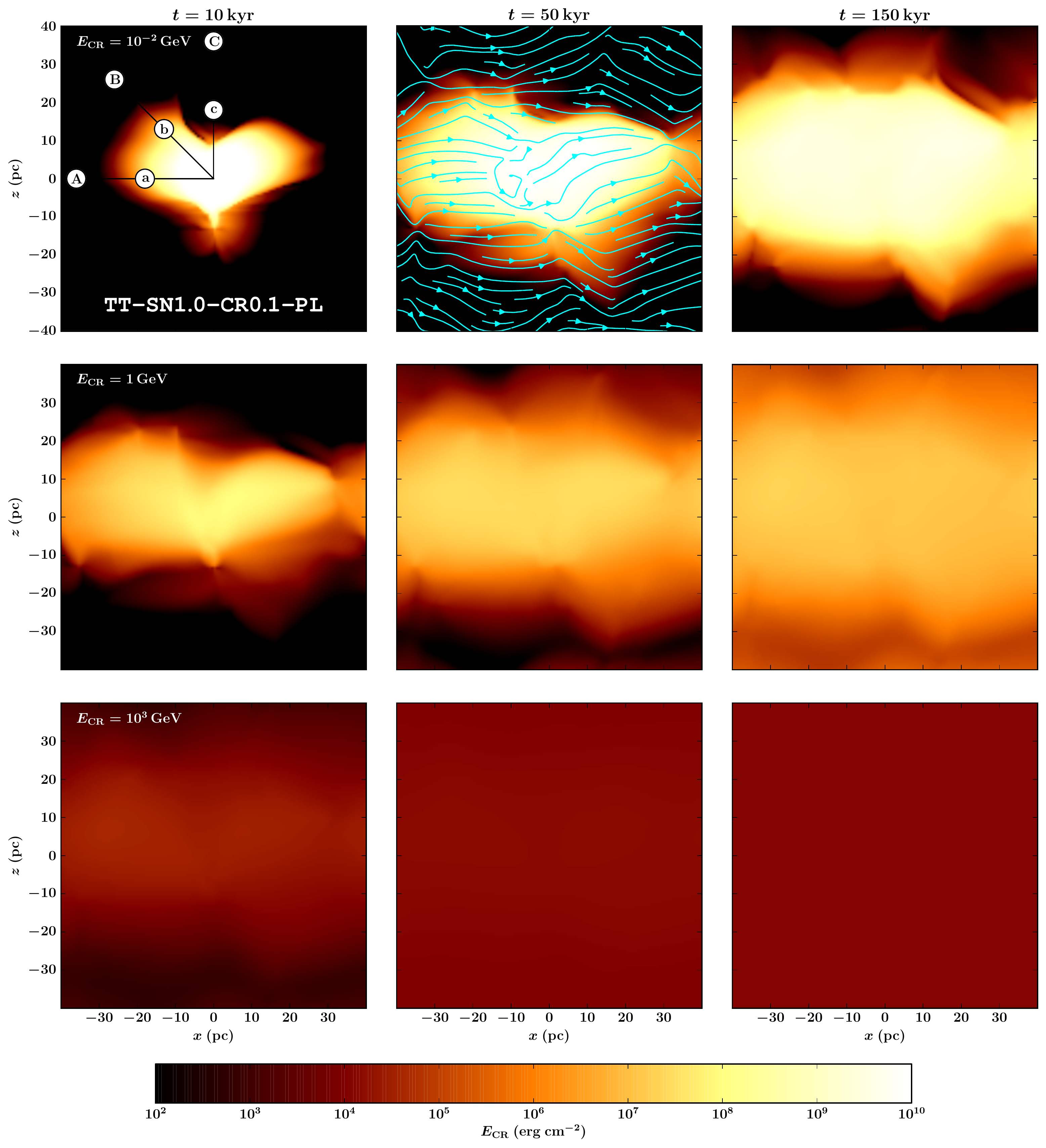}
    \includegraphics[width=16cm]{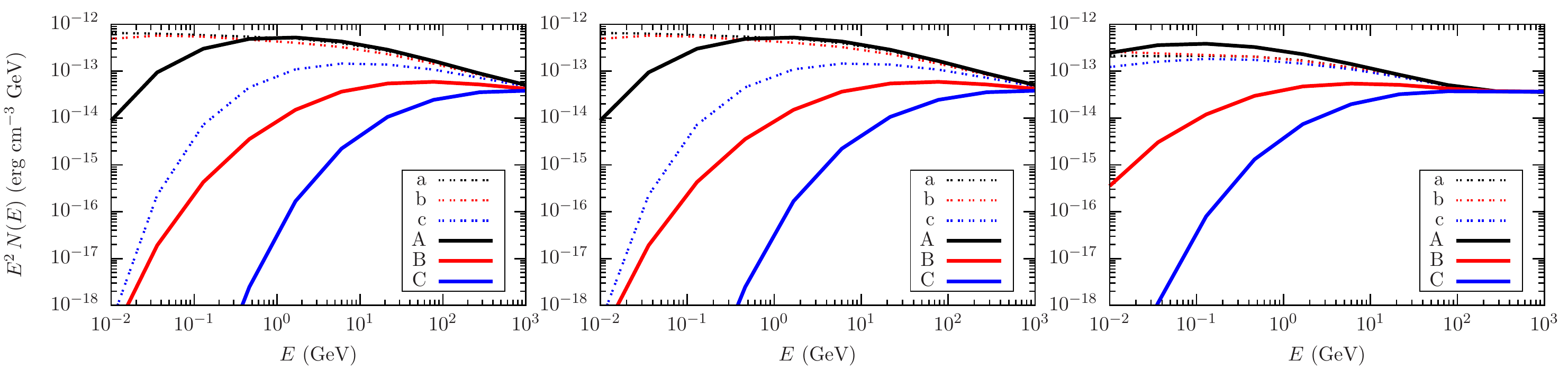}
    \caption{Projected CR energy density for different times (columns, $10\,\kyr$, $50\,\kyr$, $150\,\kyr$) and different CR energies (rows, $E_\mathrm{CR}=10^{-2}\,\GeV$, $E_\mathrm{CR}=1\,\GeV$, $E_\mathrm{CR}=10^{3}\,\GeV$) as well as the CR spectra measured at the points indicated in top left panel for simulation \texttt{TT-SN1$\!$.$\!$0-CR0$\!$.$\!$1-PL}. The cyan streamlines show the magnetic field configuration. Note that the time scale is an order of magnitude larger than the time scales in Fig.~\ref{fig:A-diff-overview} and \ref{fig:B-diff-overview}. The spectra show significant differences in total energy as well as spectral shape for the different measurement points. In this more realistic density and magnetic field configuration the measured spectrum after $t=150\,\kyr$ approaches the shape of the power-law source spectrum.}
    \label{fig:S1-B2-f1-coldens-diff-time}
  \end{minipage}
\end{figure*}

The simulations with setup TT have a self-consistently evolved density and magnetic field structure. The collapsing overdensities drag the magnetic field lines with them and increase the magnetic field energy density in dense regions. The magnetic field lines then point towards these dense regions, channelling CRs into them. Fig.~\ref{fig:S1-B2-f1-coldens-diff-time} shows the time evolution of the CR energy density for setup \texttt{TT-SN1$\!$.$\!$0-CR0$\!$.$\!$1-PL} ($B_0=2\,\mu\Gauss$, $f_\mathrm{cr}=0.1$) as well as the spectra measured at points a, b, c, and A, B, C (as in Fig.~\ref{fig:A-diff-overview}, see top left panel). The high-energy CRs diffuse through the entire box and have reached the boundary of the box after less than $10\,\kyr$. The anisotropic diffusion coefficients cause the CR energy distribution to follow the magnetic field lines reflecting the anisotropy of the magnetic field lines except for the high-energy CRs, whose fast diffusion can obliterate all directional imprints of the magnetic field. The spectra show large local differences of up to an order of magnitude. For smaller diffusion coefficients (low energy bins) the differences in the spectra are larger. At early times after the CR injection when the CRs reach the measurement points, the spectra are significantly flatter than the source spectrum. At later times, $t=150\,\kyr$, the evolved spectra approach the form of the source spectrum. For measurement point a and b the curves are almost flat, so $N(E)\propto E^{-2}$.

The projections show local enhancements of CR energy density, i.e., regions towards which the CRs diffuse more quickly. However it is important to note that the diffusion process does not accumulate energy in any particular region. Local enhancements in CR energy are not stable configurations unless a constant energy supply (from a region with even higher CR energy density) provides a steady flow of CRs. Local peaks in CR energy are caused by adiabatic compression, if the compression is faster than the diffusion process that counteracts it. However, the gas dynamics on scales of the investigated volume typically acts on longer time scales, so net compression is of less importance here.

\begin{figure}
    \centering
    \includegraphics[width=8cm]{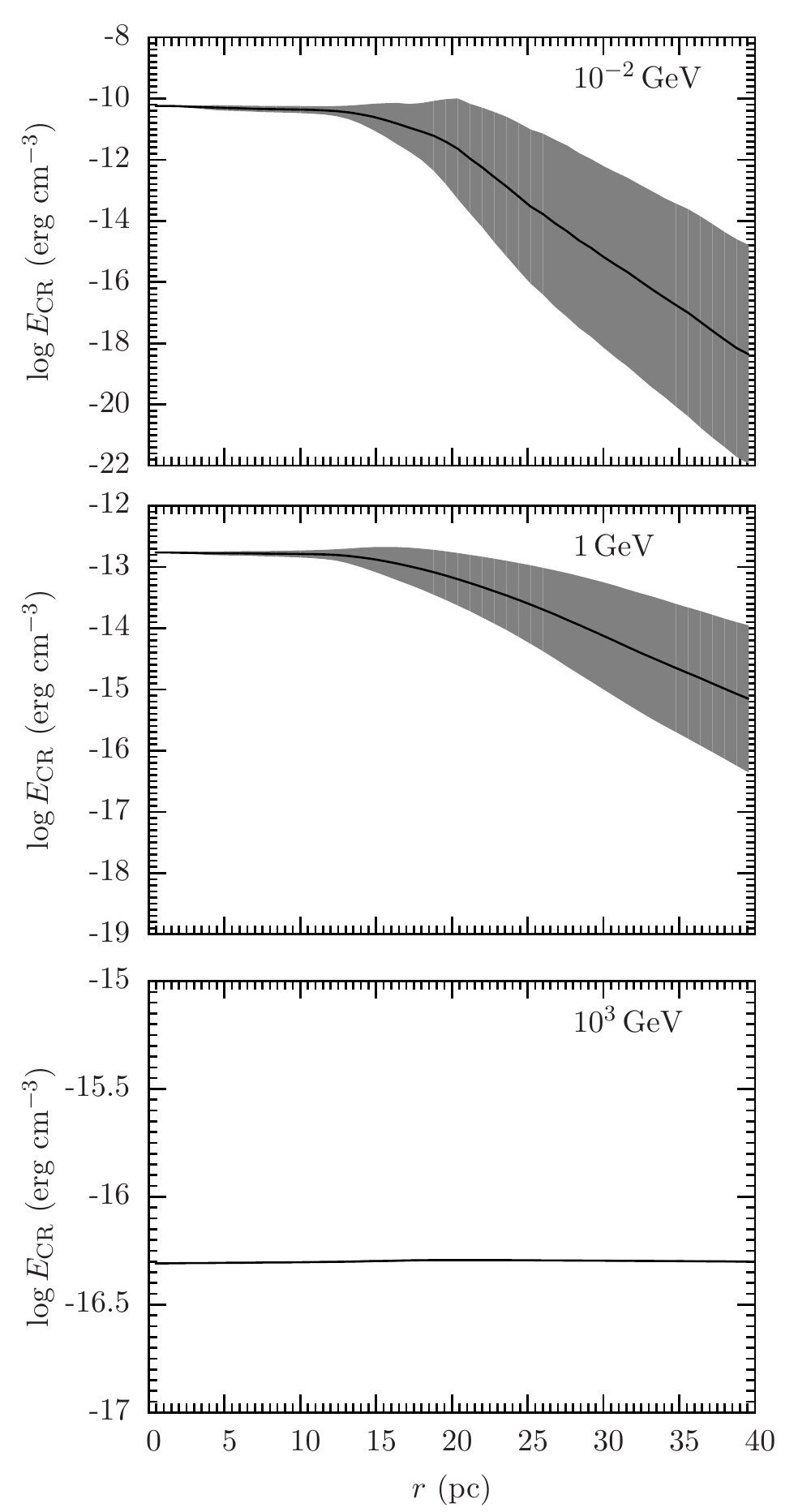}
    \caption{Radial energy distribution of the CRs for three different CR energies ($E_\mathrm{CR}=10^{-2}\,\GeV$, $E_\mathrm{CR}=1\,\GeV$, and $E_\mathrm{CR}=10^{3}\,\GeV$ at $t=100\,\kyr$) for simulation \texttt{TT-SN1$\!$.$\!$0-CR0$\!$.$\!$1-PL} (Fig.~\ref{fig:S1-B2-f1-coldens-diff-time}). The grey area indicates the angular variations computed as the arithmetic mean of the logarithm of the CR energy densities. For the low and intermediate energy CRs the angular variations are at least one order of magnitude. The high-energy CRs diffused through the box and form a uniform background CR background.}
    \label{fig:S1-B2-f1-radial-distribution}
\end{figure}

As visually suggested in the previous plots, the CR energy densities are distributed anisotropically. A more quantitative measure is depicted in Fig.~\ref{fig:S1-B2-f1-radial-distribution}, where we plot the radial distribution of CR energy density for simulation \texttt{TT-SN1$\!$.$\!$0-CR0$\!$.$\!$1-PL} at $t=100\,\kyr$ after the explosion. The top panel shows $E_\mathrm{CR}=10^{-2}\,\GeV$, the middle panel $E_\mathrm{CR}=1\,\GeV$, and the lower panel $E_\mathrm{CR}=10^{3}\,\GeV$, respectively. The grey shaded regions show the variations and indicate that for low and intermediate energy CRs the energy density can vary by an order of magnitude. For CRs with $10^3\,\GeV$ the simulated time scales are much larger than the diffusion time scale (see next section) and the CRs are uniformly distributed in the box, not reflecting the structure of the magnetic field.

\subsection{Energy transfer from CRs to gas}%

The injected CRs can interact with the gas via the additional contribution to the total pressure and the resulting pressure gradient in the MHD-CR equations. This allows for the conversion of CR energy into kinetic energy of the gas. Vice versa, adiabatic compression can convert kinetic energy into CR energy. The diffusion process counteracts this conversion by flattening the CR energy density, i.e. decreasing the CR pressure gradient. In the extreme case of infinitely large diffusion coefficients, the CR energy density will instantaneously react to any adiabatic change by flattening the CR energy to a background value. Without diffusion ($K_\parallel=K_\perp=0$) the CR fluid will react fully adiabatically. In order to estimate the net effects within this combined fluid, it is instructive to compare the characteristic time scales for diffusion and dynamical motions of the gas.

The characteristic length scale for diffusion is given by $l_\mathrm{D}=2\sqrt{Kt\,}$, which can be inverted to give a diffusion time, $t_\mathrm{D}=l_\mathrm{D}^2/4K$. Taking the maximum distance from the SN at the centre to the boundary of the box ($l_\mathrm{D}=L_\mathrm{box}/2$) and an average diffusion coefficient in the simulation of $K(E)=10^{27}(E/10\,\GeV)^{0.5}$ yields $t_\mathrm{D,CR}(E)\approx110\,\kyr\,(E/10\,\GeV)^{-0.5}$. For the low-energy CRs this is $t_\mathrm{D,CR}(E=10^{-2}\,\GeV)\approx3.4\,\Myr$, for the high energy CRs we find $t_\mathrm{D,CR}(E=10^{3}\,\GeV)\approx11\,\kyr$. The dynamical time can be estimated using the r.m.s. velocity of the gas, $v_\mathrm{rms}$, giving $t_\mathrm{hydro}=0.5L_\mathrm{box}/v_\mathrm{rms}$. The simulations show mass-weighted r.m.s. velocities ranging from $5-7\,\kmpersec$ giving characteristic time scales for the hydrodynamics of $t_\mathrm{hydro}=6-8\,\Myr$. Overall the diffusion time scales are significantly shorter than the dynamical time scales except for the low-energy CRs. Depending on the source spectrum, namely the position of the peak with most of the total CR energy, and the corresponding diffusion speeds we expect the CRs to interact very differently with the gas.

\subsection{Pressure gradients}

\begin{figure*}
  \begin{minipage}{\textwidth}
    \centering
    \includegraphics[width=16cm]{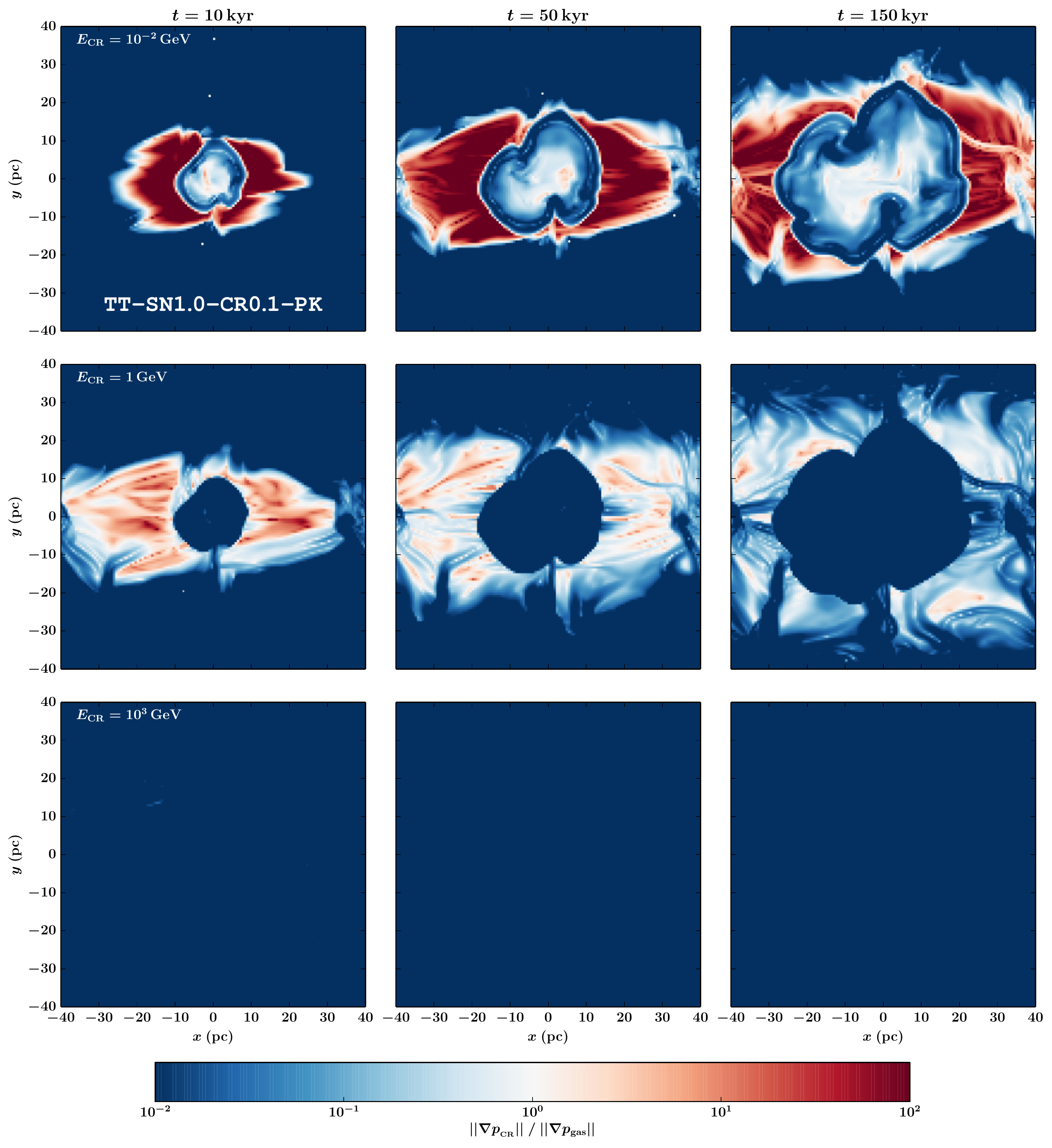}
    \caption{Ratio of the gradient of pressures, $(||\nabla \prcr||)~/~(||\nabla p_{_\mathrm{gas}}||)$, for three different times (left to right) and three different CR energies for simulation \texttt{TT-SN1$\!$.$\!$0-CR0$\!$.$\!$1-PL}. Blue regions indicate where the acceleration of the gas is dominated by gas pressure, red areas indicate CR dominant acceleration. The spatial structure of the CR dominated acceleration resembles the magnetic field structure.}
    \label{fig:S1-B2-f1-gradient-ratio}
  \end{minipage}
\end{figure*}

\begin{figure*}
  \begin{minipage}{\textwidth}
    \centering
    \includegraphics[width=16cm]{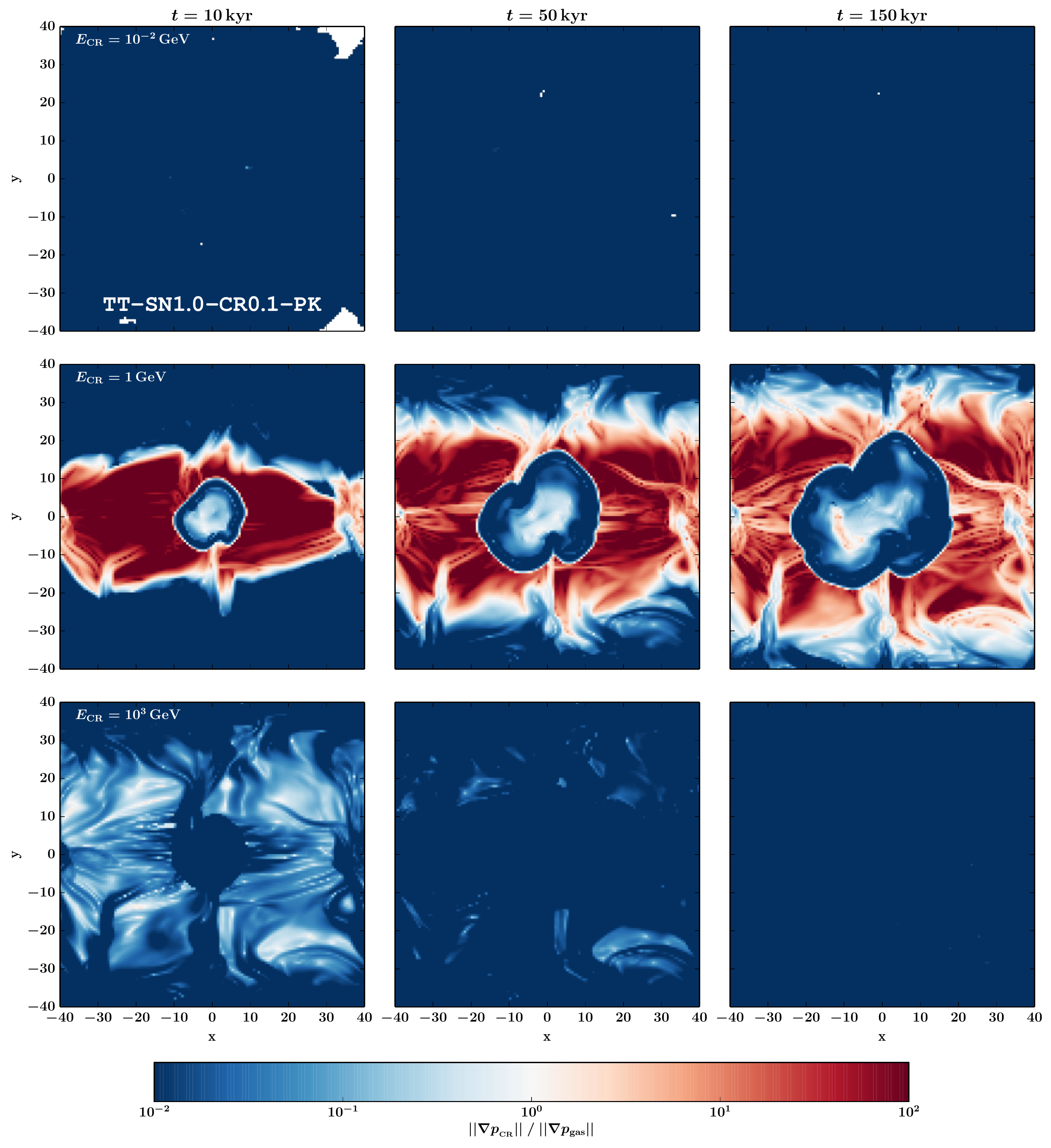}
    \caption{Ratio of the gradient of pressures, $(||\nabla \prcr||)~/~(||\nabla p_{_\mathrm{gas}}||)$, for three different times (left to right) and three different CR energies for simulation \texttt{TT-SN1$\!$.$\!$0-CR0$\!$.$\!$1-PK} (spectral peak at $E_\mathrm{CR}=1\,\GeV$). Blue regions indicate where the acceleration of the gas is dominated by gas pressure, red areas indicate CR dominant acceleration. The spatial structure of the CR dominated acceleration resembles the magnetic field structure.}
    \label{fig:S1-B2-f1-new-spec-gradient-ratio}
  \end{minipage}
\end{figure*}

In order to illustrate where the CRs are dominant in accelerating the gas, we show slices of the ratio of the CR pressure gradient to the gas pressure gradient,
\begin{equation}
  \frac{||\nabla P_\mathrm{CR}||}{||\nabla P_\mathrm{gas}||},
\end{equation}
in Fig.~\ref{fig:S1-B2-f1-gradient-ratio} (run \texttt{TT-SN1$\!$.$\!$0-CR0$\!$.$\!$1-PL} with spectral peak of the input spectrum at $E_\mathrm{CR}=10^{-2}\,\GeV$) and Fig.~\ref{fig:S1-B2-f1-new-spec-gradient-ratio} (run \texttt{TT-SN1$\!$.$\!$0-CR0$\!$.$\!$1-PK} with spectral peak of the input spectrum at $E_\mathrm{CR}=1\,\GeV$). We plot this ratio for three different times ($t=10\,\kyr$, $t=50\,\kyr$, $t=150\,\kyr$) and for three different CR energies ($E_\mathrm{EC}=10^{-2}\,\GeV$, $E_\mathrm{EC}=1\,\GeV$, $E_\mathrm{EC}=10^3\,\GeV$). Blue colours mark a negligible impact of the CRs, whereas red areas correspond to regions where CRs drive the acceleration of the gas. By comparing Fig.~\ref{fig:S1-B2-f1-gradient-ratio} with Fig.~\ref{fig:S1-B2-f1-new-spec-gradient-ratio} the impact of the input spectrum becomes evident. For the power-law input spectrum the low energy CRs contribute most to the ratio (upper panels in Fig.~\ref{fig:S1-B2-f1-gradient-ratio}). If the peak of the input spectrum is shifted to $1\,\GeV$ this becomes the most important contribution (middle panels in Fig.~\ref{fig:S1-B2-f1-new-spec-gradient-ratio}). The central blue cloud shows the expanding SN shell, where clearly gas pressure dominates over CR pressure. The diffusion coefficients for high energy CRs are very high resulting in quickly dissolving pressure gradients for this energy range.

\begin{figure*}
  \begin{minipage}{\textwidth}
    \centering
    \includegraphics[width=17cm]{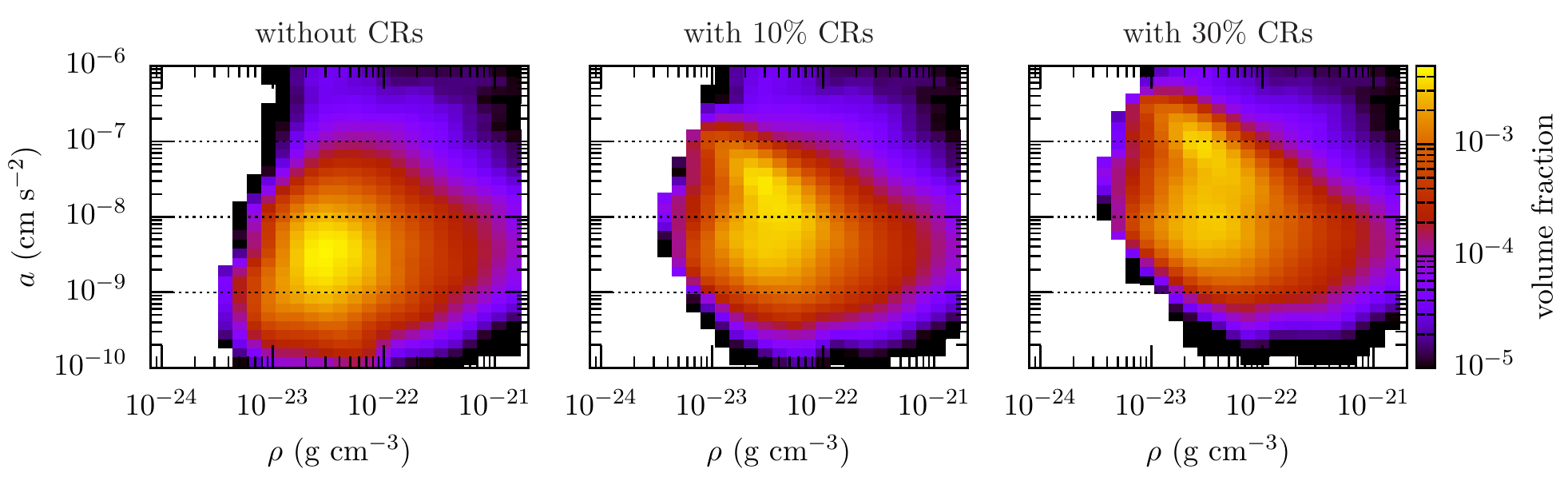}
  \end{minipage}
  \caption{Comparison of the gas acceleration without CRs (left, \texttt{TT-SN1$\!$.$\!$0-CR0$\!$.$\!$0}), with low CRs fraction (center, \texttt{TT-SN1$\!$.$\!$0-CR0$\!$.$\!$1-PK}), and high CR fraction (right, \texttt{TT-SN1$\!$.$\!$0-CR0$\!$.$\!$3-PK}) at $t=50\,\kyr$. The entire distribution is shifted to higher accelerations. Even a low fraction of CRs can dominate the acceleration of the low-density gas by increasing the values to $a\sim10^{-8}-10^{-7}\,\cm\,\second{-2}$. High-density gas is hardly affected by the CRs.}
  \label{fig:comparison-acceleration}
\end{figure*}

We quantitatively investigate the impact of CRs on the gas by analysing the net acceleration,
\begin{equation}
  \mathbf{a} = \frac{1}{\rho}\nabla P_\mathrm{tot},
\end{equation}
where $P_\mathrm{tot}$ is the total pressure, see Equation~(\ref{eq:total-pressure}). In order not to be dominated by the SN shell we exclude this region. As the ISM surrounding the SN is cold ($T\sim30\,\Kelvin$), we simply exclude the regions hotter than $10^4\,\Kelvin$. In addition the volume with negligible CR energy content ($\encr<10^{-12}\,\erg\,\cm^{-3}$) is excluded. We mask the volume based on these two criteria in the simulation with high CR fraction (\texttt{TT-SN1$\!$.$\!$0-CR0$\!$.$\!$3-PK}) and then compare the acceleration of the gas within this masked volume in three simulations in Fig.~\ref{fig:comparison-acceleration}. The left plot (\texttt{TT-SN1$\!$.$\!$0-CR0$\!$.$\!$0}) shows the distribution of accelerations resulting from gas pressure and self-gravity. The bulk of the distribution spans values of $a\sim10^{-9}-10^{-8}\,\cm\,\second^{-2}$. Once CRs are included (centre and right panel for simulations \texttt{TT-SN1$\!$.$\!$0-CR0$\!$.$\!$1-PK} and \texttt{TT-SN1$\!$.$\!$0-CR0$\!$.$\!$3-PK}) a noticeable fraction of the volume is affected, shifting the entire distribution to larger accelerations with the peak of the distribution enhanced to accelerations of $a\sim10^{-8}-10^{-7}\,\cm\,\second^{-2}$ for 10\% of CRs and values of $a\sim10^{-7}\,\cm\,\second^{-2}$ for 30\% CR input. Over time the diffusion reduces the pressure gradients and the acceleration due to CRs decreases. Nevertheless, over the entire simulation time the CRs can provide a net acceleration of $a\gtrsim10^{-8}\,\cm\,\second^{-2}$ for a significant fraction of the volume. To estimate the resulting velocities of the gas we find $v(t=150\,\kyr)=at=10^{-8}\,\cm\,\second^{-2}\,150\,\kyr\approx0.5\,\kmpersec$ as a lower limit, corresponding to a mildly supersonic flow in the cold ISM ($T\approx 30\,\Kelvin$). The net velocities in our simulations are thus not expected to increase dramatically. However, for longer integration times (and larger boxes) we note that the accumulated effect might easily provide enough support for significant acceleration of the gas, in particular if the effect of several SNe adds up to provide a coherent acceleration.

\begin{figure*}
  \begin{minipage}{\textwidth}
    \centering
    \includegraphics[width=17cm]{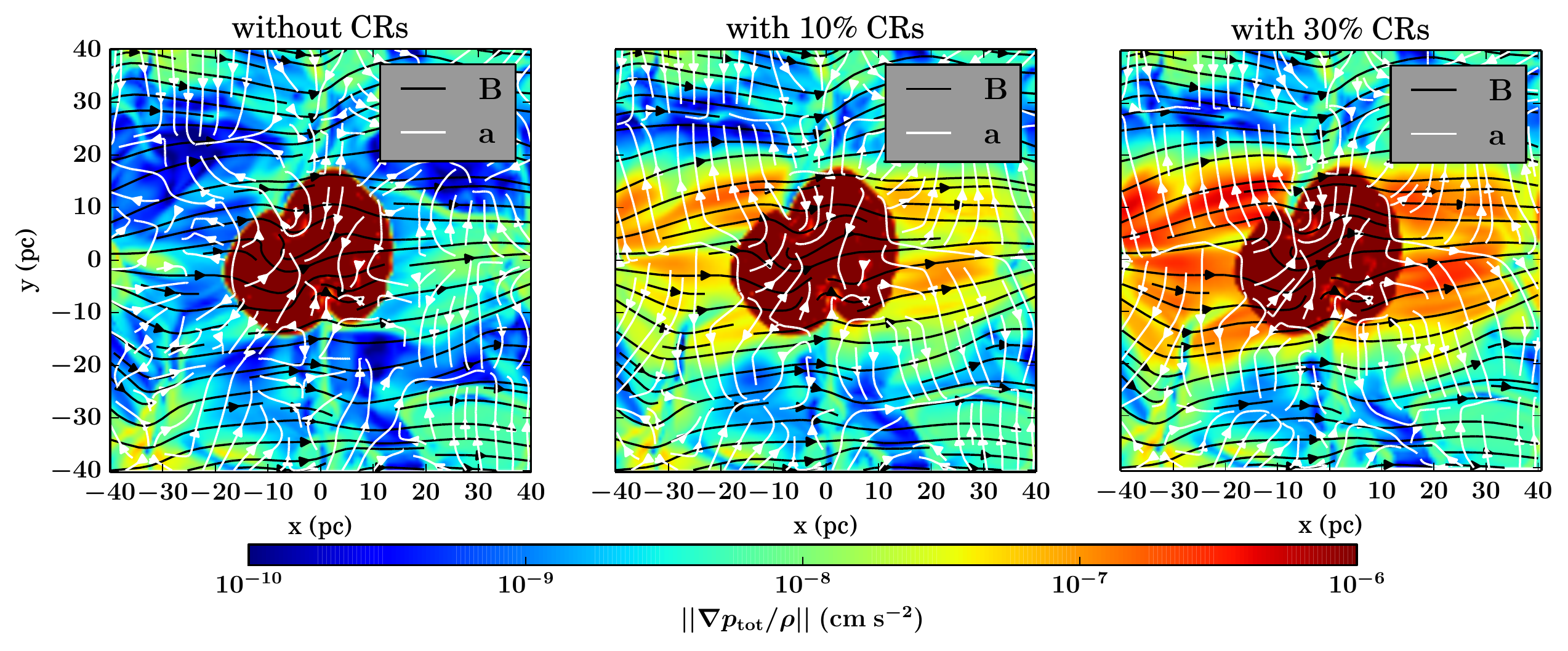}
  \end{minipage}
  \caption{Comparison of the gas acceleration without CRs (left, \texttt{TT-SN1$\!$.$\!$0-CR0$\!$.$\!$0}), with low CR fraction (center, \texttt{TT-SN1$\!$.$\!$0-CR0$\!$.$\!$1-PK}), and high CR fraction (right, \texttt{TT-SN1$\!$.$\!$0-CR0$\!$.$\!$3-PK}) at $t=50\,\kyr$. Colour-coded is the total acceleration, the black streamlines indicate the direction of the magnetic field. The white streamlines show the orientation of the acceleration, which is predominantly perpendicular to the magnetic field. Higher CR fractions increase the absolute value of the acceleration but do not change the systematic behaviour.}
  \label{fig:comparison-acceleration-streamlines}
\end{figure*}

Fig.~\ref{fig:comparison-acceleration-streamlines} further illustrates the acceleration details. Colour-coded is the modulus of the acceleration, the black streamlines show the magnetic field direction and the white streamlines the orientation of the acceleration. The left panel shows the simulation without CRs (\texttt{TT-SN1$\!$.$\!$0-CR0$\!$.$\!$0}) with accelerations $a\lesssim10^{-8}\,\cm\,\second^{-2}$. The central and right panel show the simulations including CRs (\texttt{TT-SN1$\!$.$\!$0-CR0$\!$.$\!$1-PK} and \texttt{TT-SN1$\!$.$\!$0-CR0$\!$.$\!$3-PK}). The direction of the magnetic field indicates where the CRs diffuse fastest, which is reflected in the overall higher acceleration (see also Fig.~\ref{fig:S1-B2-f1-coldens-diff-time}). The much smaller diffusion coefficients perpendicular to the field lines naturally causes the CR pressure gradient to be larger perpendicular to the field lines. It is thus not surprising that in the runs with CRs the acceleration of the gas due to CRs is mainly perpendicular to the magnetic field lines. This is illustrated particularly well on the left-hand side of the SN remnant. The previously randomly oriented acceleration vectors are much stronger oriented with respect to the magnetic field. A higher CR fraction increases the net accelerations but does not change the systematic behaviour.

Both the distribution of the acceleration (Fig.~\ref{fig:comparison-acceleration}) as well as the slice plots with the direction of the field vectors (Fig.~\ref{fig:comparison-acceleration-streamlines}) show that the CRs can efficiently contribute to the local acceleration of the gas. This indicates that CRs can significantly contribute to the gas motions in regions unaffected by the expanding SN remnant.

\subsection{Velocity dispersion and kinetic energy}%

The integrated effect of the CR acceleration is expected to be small on the timescales considered here, if just considering the gas ahead of the SN shell. However, the SN shell itself will also be affected by the CR--gas interaction.

\begin{figure}
  \centering
  \includegraphics[width=8cm]{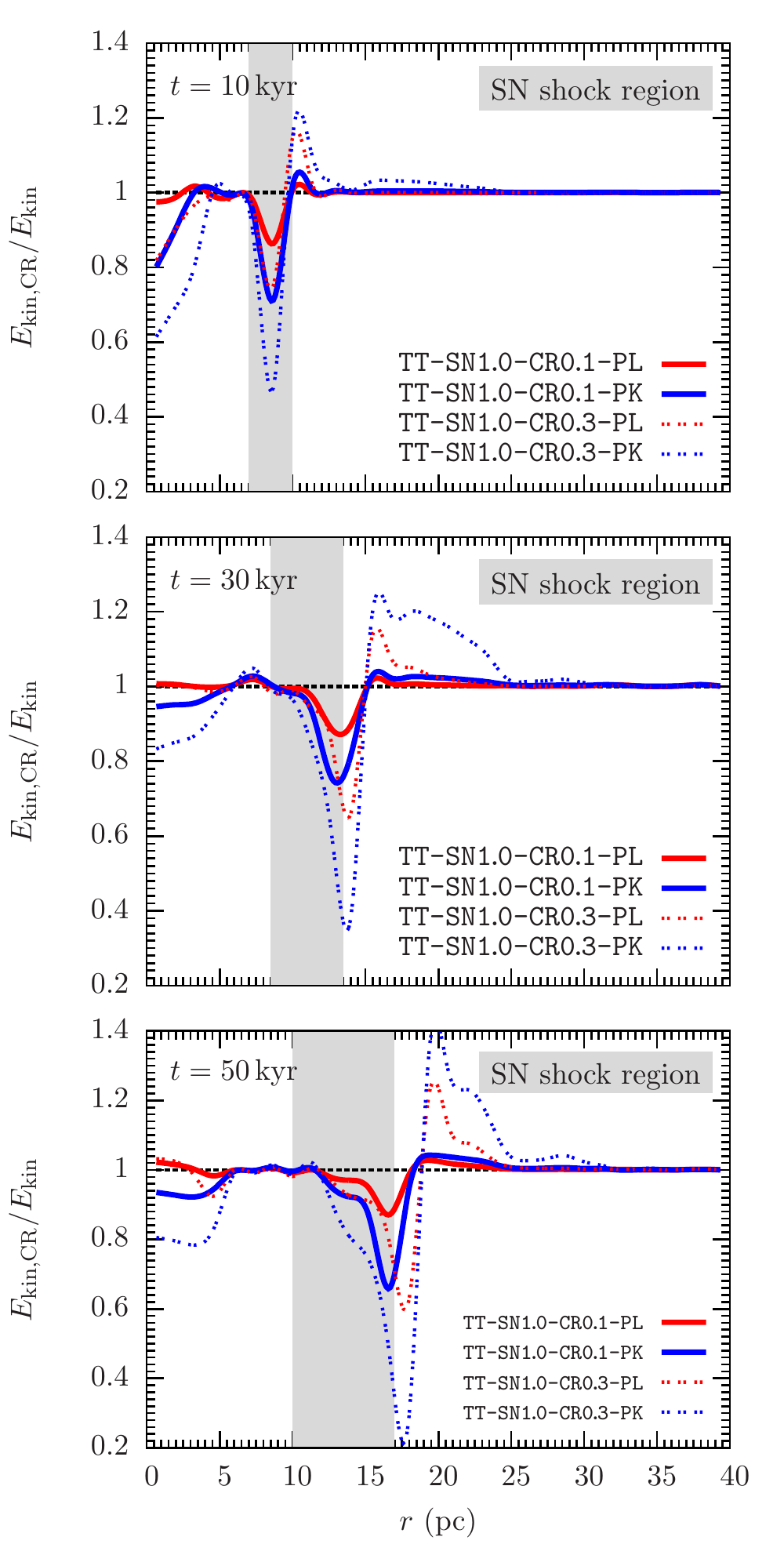}
  \caption{Kinetic energy as a function of radius compared to the fiducial run \texttt{TT-SN1$\!$.$\!$3-CR0$\!$.$\!$0-PL} without CRs (black lines). The grey area indicates the radial extent of the SN shell. The red curve (\texttt{TT-SN1$\!$.$\!$0-CR0$\!$.$\!$3-PL}) shows the run with 30\% of the total SN energy in CRs and a simple CR source spectrum, $N(E)\propto E^{-2}$. The blue line corresponds to \texttt{TT-SN1$\!$.$\!$0-CR0$\!$.$\!$3-PK-ad} with the modified CR source spectrum peaking at $1\,\GeV$ (equation~\ref{eq:source-spectrum2}). The CRs take energy out of the immediate shock region and transport it to larger distances ahead of the SN shell. This effect is stronger for a peaked input spectrum.}
  \label{fig:S1-ekin}
\end{figure}

The net impact of the CRs on the acceleration of the gas is depicted in Fig.~\ref{fig:S1-ekin}. The plots show the radial distribution of the kinetic energy around the site of the SN, averaged over the $4\pi$ solid angle. We compare the simulations with CRs to the corresponding simulation with the same total energy but only injected as thermal energy, namely simulations \texttt{TT-SN1$\!$.$\!$0-CR0$\!$.$\!$1-PL} and \texttt{TT-SN1$\!$.$\!$0-CR0$\!$.$\!$1-PK} to setup \texttt{TT-SN1$\!$.$\!$1-CR0$\!$.$\!$0} for a low fraction of CRs, as well as simulations \texttt{TT-SN1$\!$.$\!$0-CR0$\!$.$\!$3-PL} and \texttt{TT-SN1$\!$.$\!$0-CR0$\!$.$\!$3-PK} to setup \texttt{TT-SN1$\!$.$\!$3-CR0$\!$.$\!$0} for a high fraction of CRs. We do not include the corresponding runs with adiabatic losses because they differ by a negligible amount. The runs including CRs show a systematic trend compared to the purely thermal run. The kinetic energy at the SN shock front (indicated by the grey area in the plot) is reduced compared to the run without CRs. This effect scales with the total fraction of CRs. The CRs then diffuse ahead of the shock front and can accelerate the gas at larger radii. However, by how much the gas is accelerated immediately ahead of the shock front does not scale linearly with the fraction of CR input. For the low CR fraction the acceleration ahead of the shock front is significantly smaller ($\sim 5\%$) compared to the runs with higher CR fraction ($\sim20-40\%$). Over time, this effect accumulates until the CRs diffused out of the box and numerical dissipation dominates.

The overall efficiency of CR energy conversion is very difficult to determine with our setup. A global comparison of how much the kinetic energy is enlarged by the presence of CR gives efficiencies at a percent level. However, with outflow boundaries it is difficult to precisely determine how much energy left or entered the box. Applying periodic boundary conditions would solve the problem of the unknown fraction of diffused CRs, but would strongly affect the CR energy distribution because of a CR energy density background and thus the CR pressure gradients and the conversion efficiency. In principle we can integrate the total amount of CR energy that leaves the box and account for the turbulent energy that enters and leaves the box. But the overall efficiency of CR energy conversion is better studied statistically in larger boxes with numerous SNe, where precision measurements are not influenced by statistical noise of the local magnetic field configuration or the density distribution.

\begin{figure}
    \centering
    \includegraphics[width=8cm]{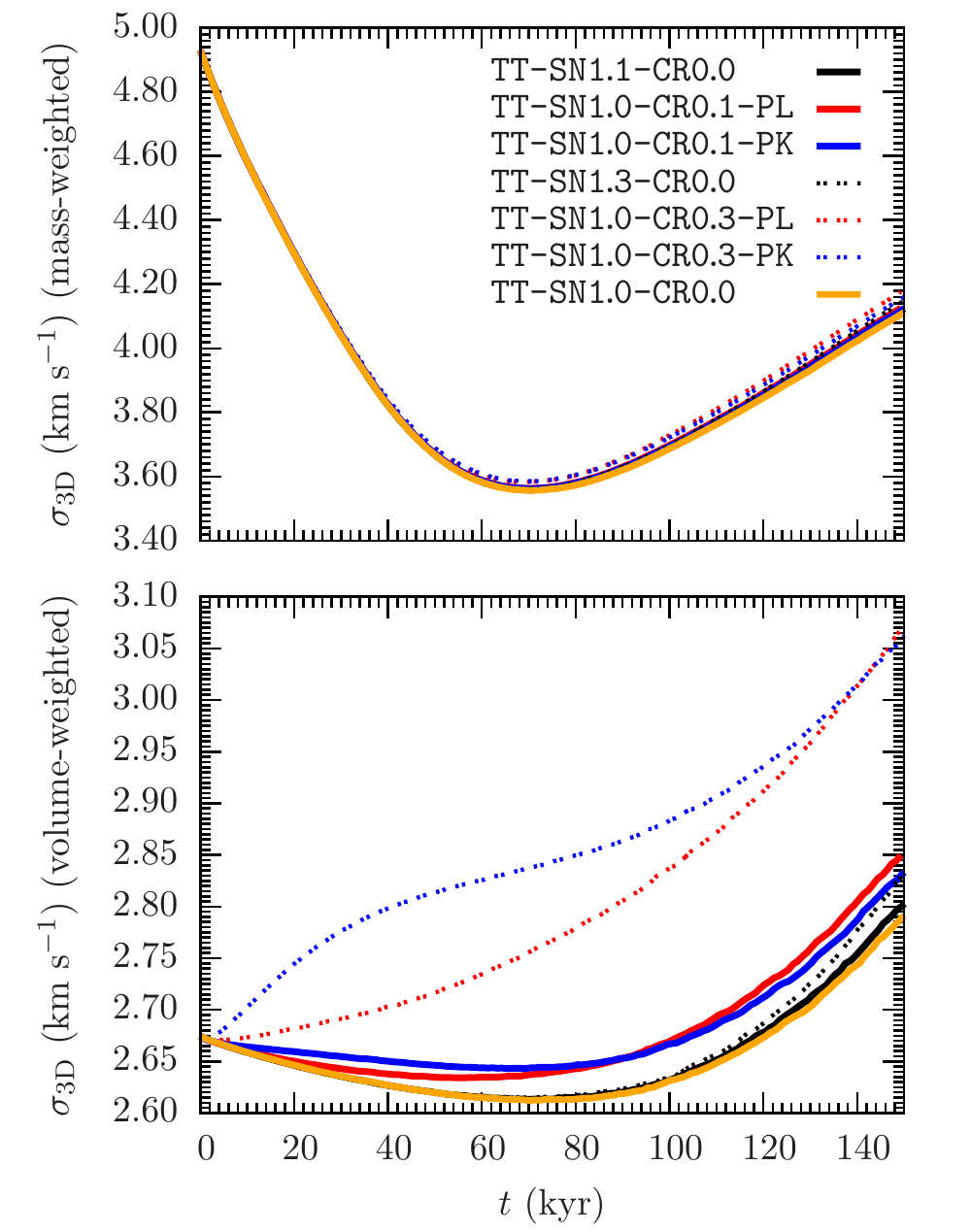}
    \caption{Velocity dispersion of the gas excluding the region of the SN remnant, separately plotted for volume and mass-weighted velocity dispersion as a function of time. The dense regions are unaffected by the CR acceleration, which is reflected in the mass-weighted curves. Low-density regions are accelerated to higher velocities with a strongly non-linear dependence on the CR fraction $f_\mathrm{CR}$.}
    \label{fig:velocity-dispersion}
\end{figure}

A simple measurable quantity that indicates the net dynamical impact of the CRs is the three-dimensional velocity dispersion,
\begin{equation}
  \sigma_\mathrm{3D} = \sqrt{\sum_j \sigma_j^2},
\end{equation}
with $j\in [x, y, z]$. The mass-weighted one-dimensional component is given by 
\begin{equation}
  \sigma_j = \sqrt{\frac{1}{M_\mathrm{tot}}\sum_\mathrm{cells}\rkl{m(v_j-\skl{v}_j)^2}},
\end{equation}
where $\skl{v}_j$ is the mass-weighted average velocity in one direction,
\begin{equation}
  \skl{v}_j = \frac{1}{M_\mathrm{tot}}\sum_\mathrm{cells}\rkl{mv_j}.
\end{equation}
In this analysis, we exclude the SN shell (gas with $T>10^4\,\Kelvin$) as well as the regions with negligible CR energy content ($\encr<10^{-12}\,\erg\,\cm^{-3}$).
We also compute the corresponding volume-weighted values, both shown in Fig.~\ref{fig:velocity-dispersion} as a function of time after the SN explosion for simulations \texttt{TT-SN1$\!$.$\!$0-CR0$\!$.$\!$1-PL} and \texttt{TT-SN1$\!$.$\!$0-CR0$\!$.$\!$1-PK} compared to \texttt{TT-SN1$\!$.$\!$1-CR0$\!$.$\!$0} as well as the corresponding set of simulations with hier CR fraction (\texttt{TT-SN1$\!$.$\!$0-CR0$\!$.$\!$3-PL} and \texttt{TT-SN1$\!$.$\!$0-CR0$\!$.$\!$3-PK} compared to \texttt{TT-SN1$\!$.$\!$3-CR0$\!$.$\!$0}). The main shape of the curves reflects the overall decay of turbulence in first half of the simulation and the beginning collapse of one region in the second half of the simulation. The mass-weighted velocity dispersions do not show a noticeable difference. This is consistent with our previous estimates and findings of the acceleration. The volume-weighted velocity dispersion shows slightly higher values if CRs are included. The effect is relatively small at a percent level for a CR fraction of $f_\mathrm{CR}=0.1$. For the large CR fraction ($f_\mathrm{CR}=0.3$) the volume-weighted velocity dispersion increases early on in the simulation. The effect does not scale linearly with the input fraction $f_\mathrm{CR}$.

\subsection{Adiabatic losses and spectral changes}

\begin{figure}
  \centering
  \includegraphics[width=8cm]{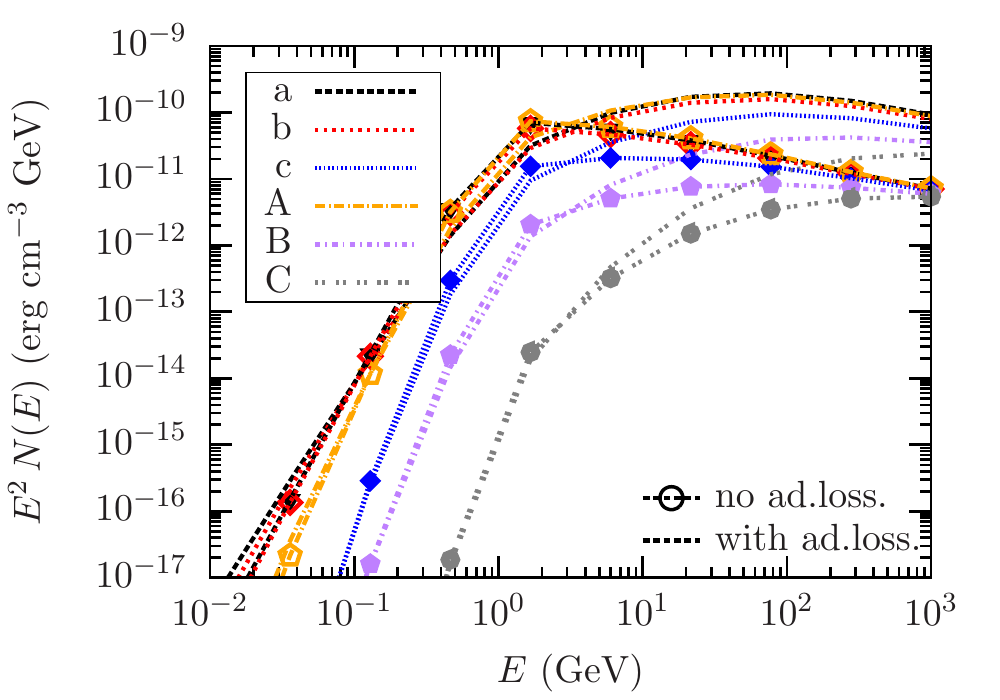}
  \caption{Spectra for \texttt{TT-SN1$\!$.$\!$0-CR0$\!$.$\!$1-PK} (lines with points) and \texttt{TT-SN1$\!$.$\!$0-CR0$\!$.$\!$1-PK-ad} (lines) at time $t=50\,\kyr$ measured at the six different measurement points indicated in Fig.~\ref{fig:A-diff-overview}. The spectra for the run including adiabatic losses show a higher total energy in the cells and are slightly harder.}
  \label{fig:spectra-comp-ad-noad}
\end{figure}

The simulation setup and the time scale emphasise the diffusion of CRs, rather than the dynamical evolution of the gas. Adiabatic losses, which are solely due to compression and expansion of the fluid, are therefore expected to play a minor role in this context. We observe significant spectral changes inside the injection region of the SN, where the reverse shock causes a hardening of the spectrum. However, our CR injection model is highly idealised. We inject CR energy in the entire injection region of the SN, whereas CRs are expected to be accelerated in the shock of the SN shell. We therefore should not investigate this region. The expanding SN shell also pushes CRs with a slightly hardened spectrum into the surrounding ISM. The reason is the compressing flow at the shock front. This behaviour is shown in Fig.~\ref{fig:spectra-comp-ad-noad}, where we plot the spectra at $t=50\,\kyr$ for simulation \texttt{TT-SN1$\!$.$\!$0-CR0$\!$.$\!$3-PK} (lines with points) and \texttt{TT-SN1$\!$.$\!$0-CR0$\!$.$\!$3-PK-ad} (lines), shown at the six measurement points indicated in Fig.~\ref{fig:A-diff-overview}. Again, this hardening proves that the code is solving the adiabatic losses correctly but the physical importance suffers from the idealised simulation setup. Ideally, the shock front would be resolved well enough to inject CRs directly with a spectrum depending on the shock properties. As the hardening of the spectrum allows more CRs to diffuse faster and reduce the CR pressure gradient faster, this effect of adiabatic losses tends to (if at all) reduce the dynamical impact of CRs.

Whereas the spectra differ noticeably in the runs with and without adiabatic losses, the overall integrated effect is rather small, because the CR pressure gradient does not change perceptibly. The differences in the kinetic energy between runs with and without adiabatic losses are less than 10\%.

\section{Summary and Discussion}

We present a numerical implementation to follow the transport of CRs and their dynamical coupling to the gas (in the advection-diffusion approximation) into the AMR-MHD code \textsc{FLASH4}. The CRs with different energies are treated as separate fluids. We include energy dependent anisotropic diffusion with respect to the direction of the magnetic field as well as adiabatic losses to follow the evolution of the CR spectra. As a first application we investigate the impact of CRs on the ISM in the vicinity of a young SN remnant in a cubic box of $(80\,\pc)^3$. We study the effect of different CR energies and different CR injection spectra. The SN explodes at the centre of the box and acts as a source of CRs. The fraction of CR energy input is varied from $0-30\%$ of the thermal SN energy. With our numerical experiment we conclude the following.

\begin{itemize}
\item CRs quickly diffuse along the magnetic field lines because the parallel diffusion coefficient is two orders of magnitude larger than the perpendicular one. Depending on the orientation of the magnetic field and the energy of the CRs the spatial distribution differs perceptibly. For realistic magnetic field configurations low and intermediate energy CRs ($E_\mathrm{CR}\lesssim1\,\GeV$) show angular anisotropies of an order of magnitude throughout the simulation ($t_\mathrm{end}=150\,\kyr$). For CRs with higher energies the diffusion time scale is shorter than the simulation time, giving a much more flattened energy distribution. The highest energy CRs ($E_\mathrm{CR}=10^3\,\GeV$) are basically uniformly distributed after $100\,\kyr$.
\item As the diffusion time scales are short in comparison to the hydrodynamical time scale, the CRs can escape the injection region and penetrate ahead of the SN shell into the surrounding ISM, where they efficiently accelerate the gas. Regions of high gas density are dominated by thermal and magnetic pressure gradients. The net acceleration of low-density regions unaffected by the SN shell is about one-two orders of magnitude larger for the runs including CRs with the acceleration mainly acting perpendicular to the magnetic field lines.
\item The integrated effect in terms of kinetic energy and velocity dispersion are rather small on the simulated spatial and temporal scales. At distances of $\sim5\,\pc$ ahead of the SN shell, the kinetic energy increases locally by $5-40\%$ compared to a setup with the same total energy but only thermal SN energy injection, depending on the source spectrum and the amount of CR energy we inject. The changes of the velocity dispersions are at the level of a few percent unless a large fraction of energy is put into CRs ($f_\mathrm{CR}=0.3$).
\item Adiabatic losses change the shape of the spectrum. However, on the simulated time scales, the spectral changes due to adiabatic losses are not significant enough to perceptibly change the total dynamical impact of the CRs on the gas. The integrated amount of the energy converted from CR energy to kinetic energy differs by only a few percent between the runs with and without adiabatic losses. Changing the shape of the source spectrum ($N(E)\propto E^{-2}$ versus peaked spectrum at $E_\mathrm{CR}=1\,\GeV$) has a much stronger effect than adiabatic losses during the evolution of the supernova remnant.
\end{itemize}

Uncertainties of the model presented in this study are the actual values of the diffusion coefficients including the ratio of the parallel to the perpendicular components and their dependence on the CR energy. In addition, we do not have loss processes for the CRs and treat the gas adiabatically. We also assume a homogeneous distribution of CRs in the entire SN injection region instead of following the CR acceleration in shocks explicitely. Finally, the total amount of CRs accelerated per SN is uncertain, which we vary from $0-30\%$ of the thermal SN energy.

Previous studies on galactic scales and much longer time scales indicate that CRs can have a significant impact on the gas dynamics in galaxies, e.g. supporting large-scale gas outflows. However, many of these studies neglect the effect of magnetic fields assuming isotropic diffusion \citep{UhligEtAl2012,BoothEtAl2013,SalemBryan2013}. Models taking anisotropic diffusion into account \citep{YangEtAl2012,HanaszEtAl2013} have so far neglected the energy dependence of the CRs as well as adiabatic losses. Our results indicate that the shape of the source spectrum in combination with adiabatic losses might have a significant on where and how CRs efficiently interact with the gas in the regions ahead of the SNe. Future simulations of the ISM on larger scales (up to a few hundred pc) taking multiple SNe into account will help to understand the CR impact, the equilibrium multiphase structure of the ISM and how the launching of Galactic winds can be supported.

\clearpage
\section*{Acknowledgements}
We thank Torsten En{\ss}lin, Ewald M\"{u}ller, Harald Lesch, Jerry Ostriker, and Damiano Caprioli for inspiring discussions. We thank Christian Karch for the program package \textsc{fy} and the community of the \textsc{yt-project} for the \textsc{yt} package \citep{TurkEtAl2011}, which we use to plot and analyse most of the data.
P.G., T.N., and S.W. acknowledge support from the DFG Priority Program 1573 {\em Physics of the Interstellar Medium}.
M.H. acknowledges a kind hospitality of the University Observatory Munich under financial support from DAAD grant
and a partial support by Polish Ministry of Science and Higher Education through the grant N203 511038.
S.W. acknowledges the support of the Bonn-Cologne Graduate School, which is funded through the Excellence Initiative.

\begin{appendix}
\label{sec:appendix}

\section{Algorithms}%
\label{sec:CR-implementation}

\subsection{Numerical treatment of CR in MHD simulations}%

Equation~\ref{eq:CR-adv-diff-glob} can be written in the conservative form
\begin{equation}
  \delt\encr + \nabla\cdot\vecF_{_\mathrm{CR,adv}} + \nabla\cdot\vecF_{_\mathrm{CR,diff}} = -\prcr\nabla\cdot\vecv + \Qcr.
\end{equation}
The terms $-\prcr\nabla\cdot\vecv$ and $\Qcr$ are source terms, $\vecF_{_\mathrm{CR,adv}} = \encr\vecv$ is the CR flux advected with the gas flow, $\vecF_{_\mathrm{CR,diff}} = -\tenK\nabla\encr$ is the diffusive flux.

We solve the MHD equations using cell centred quantities for both the CR energy as well as the magnetic field quantities, $\mathbf{B}$. This approach is different from the numerical scheme used in other codes, \citep[e.g.,][]{HanaszLesch2003}, where they use staggered mesh with the magnetic field components defined at face centres. As the fluxes are computed at the cell boundaries but the magnetic field values at the cell centres, we need to interpolate. The values at the cell boundaries are computed as discrete differences with an MC-limiter (monotonised central limiter) for oscillation control, see~\citet{Waagan2009}.

\subsection{Anisotropic diffusion}%

The diffusion of CR shows a strong dependence on the direction of the magnetic field and needs to be treated in an anisotropic way parallel and perpendicular to the magnetic field lines. The diffusion tensor $\tenK$ thus depends on the variable magnetic field configuration. We now focus on the diffusion term of the advection-diffusion equation
\begin{equation}
  \delt\encr + \nabla\cdot\vecF_{_\mathrm{CR}} = 0, \qquad \vecF_{_\mathrm{CR}} = -\tenK\nabla\encr.
\end{equation}
Discretised, the complete three-dimensional conservation law reads
\begin{align}
  e_\mathrm{CR,i,j,k}^{n+1} = e_\mathrm{CR,i,j,k}^{n}
  & - \frac{\Delta t}{\Delta x} \rkl{F_\mathrm{CR,i+\frac{1}{2},j,k} - F_\mathrm{CR,i-\frac{1}{2},j,k}}\\
  & - \frac{\Delta t}{\Delta y} \rkl{F_\mathrm{CR,i,j+\frac{1}{2},k} - F_\mathrm{CR,i,j-\frac{1}{2},k}}\\
  & - \frac{\Delta t}{\Delta z} \rkl{F_\mathrm{CR,i,j,k+\frac{1}{2}} - F_\mathrm{CR,i,j,k-\frac{1}{2}}}.
\end{align}
The quantities $e_\mathrm{CR,i,j,k}^{n+1}$ and $e_\mathrm{CR,i,j,k}^{n}$ are cell centered CR energy densities in cell $i, j, k$ at time steps $t^{n+1}$ and $t^n$, and $F_\mathrm{CR,i+\frac{1}{2},j,k}, F_\mathrm{CR,i-\frac{1}{2},j,k}$ are the fluxes of CR through the left and right boundaries of the cell in $x$ direction.

We can now combine the diffusion of CRs in a directionally split scheme to compute the total change in CR energy density. We subsequently apply the energy fluxes in $x,y,$ and $z$ direction
\begin{align}
  e_\mathrm{CR,i,j,k}^{n+b} &= e_\mathrm{CR,i,j,k}^{n+a} - \frac{\Delta t}{\Delta x} \rkl{F_\mathrm{CR,i+\frac{1}{2},j,k} - F_\mathrm{CR,i-\frac{1}{2},j,k}}\\
  e_\mathrm{CR,i,j,k}^{n+c} &= e_\mathrm{CR,i,j,k}^{n+b} - \frac{\Delta t}{\Delta x} \rkl{F_\mathrm{CR,i,j+\frac{1}{2},k} - F_\mathrm{CR,i,j-\frac{1}{2},k}}\\
  e_\mathrm{CR,i,j,k}^{n+1} &= e_\mathrm{CR,i,j,k}^{n+c} - \frac{\Delta t}{\Delta x} \rkl{F_\mathrm{CR,i,j,k+\frac{1}{2}} - F_\mathrm{CR,i,j,k-\frac{1}{2}}}
\end{align}
with $e_\mathrm{CR,i,j,k}^{n+a}$ being the CR energy density after the advection step.

\subsection{Time step limitations}%
\label{sec:time-step}

As we are using an explicit scheme for the diffusion we have to obey stability criteria for the numerical scheme. The time step limitations for the diffusion part of the CR advection-diffusion equation is
\begin{equation}
  \Delta t = 0.5\,\mathrm{CFL}_\mathrm{CR}\,\frac{\min(\Delta x, \Delta y, \Delta z)^2}{K_\parallel + K_\perp},
\end{equation}
where $\mathrm{CFL}_\mathrm{CR}$ is the Courrant-Friedrichs-Lewi number for the CR diffusion scheme and $\Delta x$, $\Delta y$, and $\Delta z$ are the cell sizes. As the diffusion equation is a second-order differential equation the limit on the time step scales as $\Delta x^2$.

\section{Test problems}%
\label{sec:CR-tests}

\subsection{Advection}

\begin{figure}
  \centering
  \includegraphics[width=8cm]{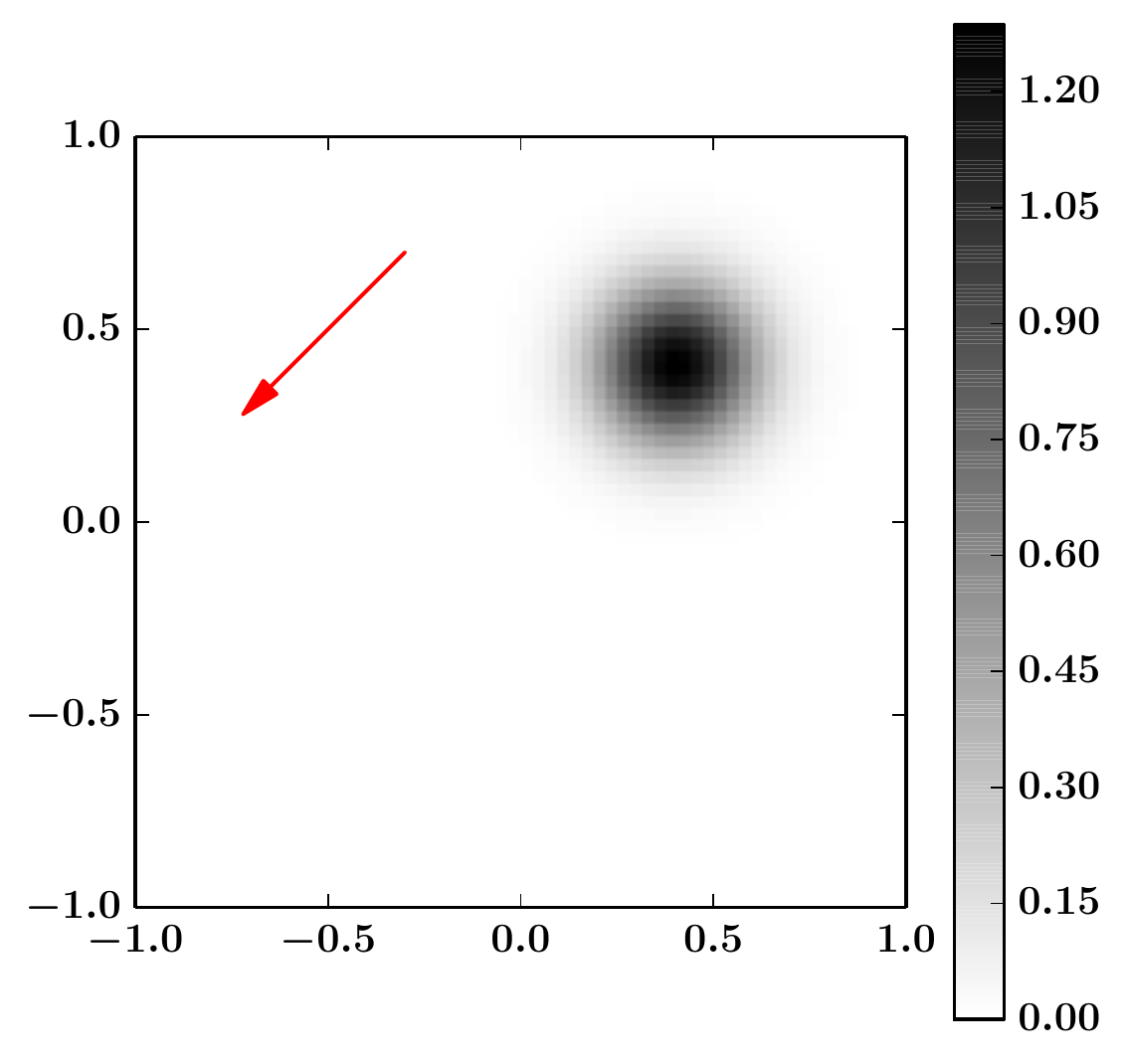}
  \includegraphics[width=8cm]{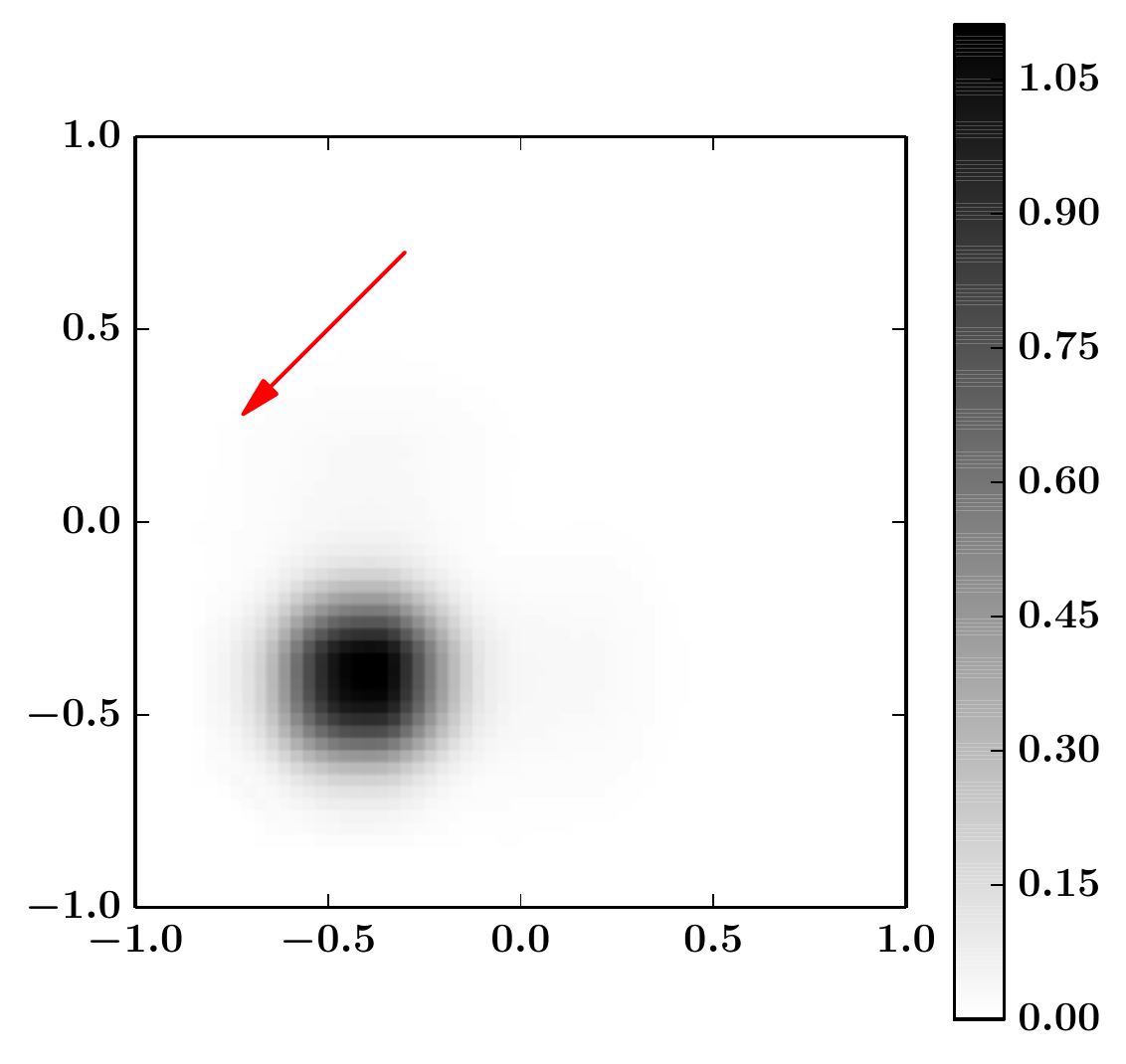}
  \caption{Advection test for the MHD-CR solver. The arrow shows the direction of the velocity, the color shows a slice of the CR energy density. \textsc{Top}: Initial CR energy density. \textsc{Bottom}: Advected CR energy density. The diffusive character of the solver results in a slightly smaller peak value of the energy density, which can be seen at the scale of the colour bar. The explicit diffusion terms are switched off.}
  \label{fig:test-advection}
\end{figure}

For this test we switch off the diffusion of the solver and let the overdensity of cosmic ray energy move with the gas flow. The CR energy is plotted for two different times in figure~\ref{fig:test-advection}. The diffusion of the CR energy density is due to the numerical diffusion of the hydro solver.

\subsection{Passive CR diffusion}%

\begin{figure}
  \centering
  \includegraphics[width=8cm]{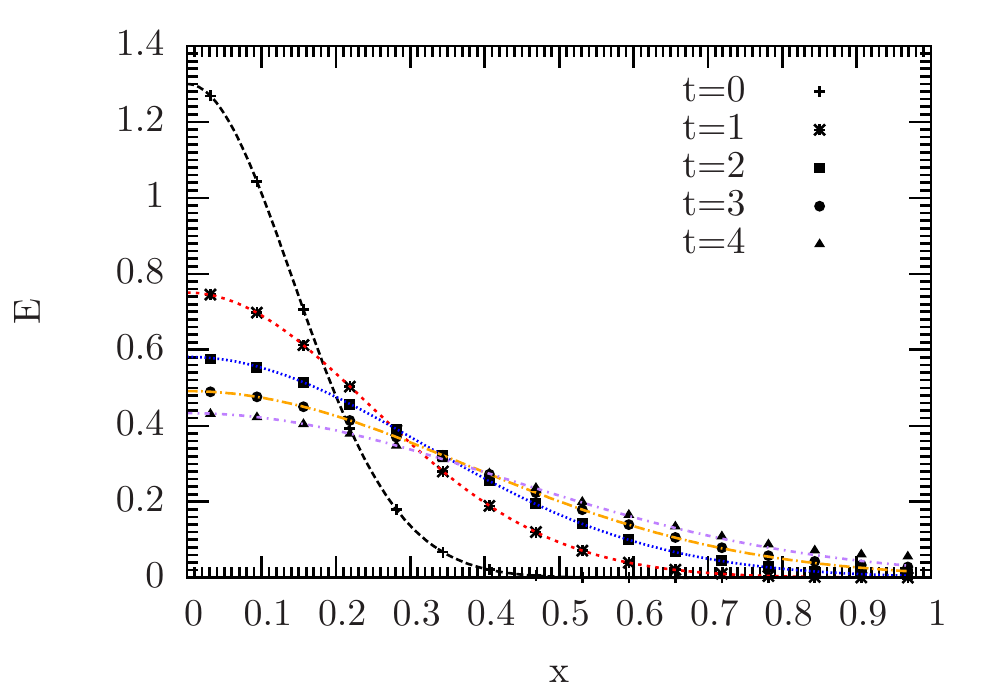}
  \caption{Diffusion test in one dimension. Shown is the analytic solution (lines) as well as the numerical values (points).}
  \label{fig:test-diffusion-1D}
\end{figure}

The most simple test is to switch off the advection of the CR fluid and compute the diffusion in one direction. Passive transport means that the CR energy density does not have a dynamical influence on the gas, i.e., the term $-\nabla\prcr/\rho$ is neglected in the equation of motion of the gas. The diffusion problem can then be reduced to the equation
\begin{equation}
  \delt\encr = K\delx^2\encr.
\end{equation}
Given an initial distribution of the CR energy of the form
\begin{equation}
  e_{_{\mathrm{CR},0}} = A\,\exp\rkl{-\frac{x^2}{r_0^2}}
\end{equation}
with an initial half-width radius, $r_0$ and an amplitude, $A$, yields the following analytic time evolution of the profile
\begin{equation}
  \encr(x,t) = A\,\sqrt{\frac{r_0^2}{r_0^2 + 4Kt}\,}\,\exp\rkl{-\frac{x^2}{r_0^2 + 4Kt}}.
\end{equation}
In figure~\ref{fig:test-diffusion-1D} we show the numerical diffusion which is in good agreement with the analytical solution.

\subsection{Anisotropic diffusion in 3D}%

\begin{figure*}
  \begin{minipage}{\textwidth}
  \centering
  \includegraphics[width=8cm]{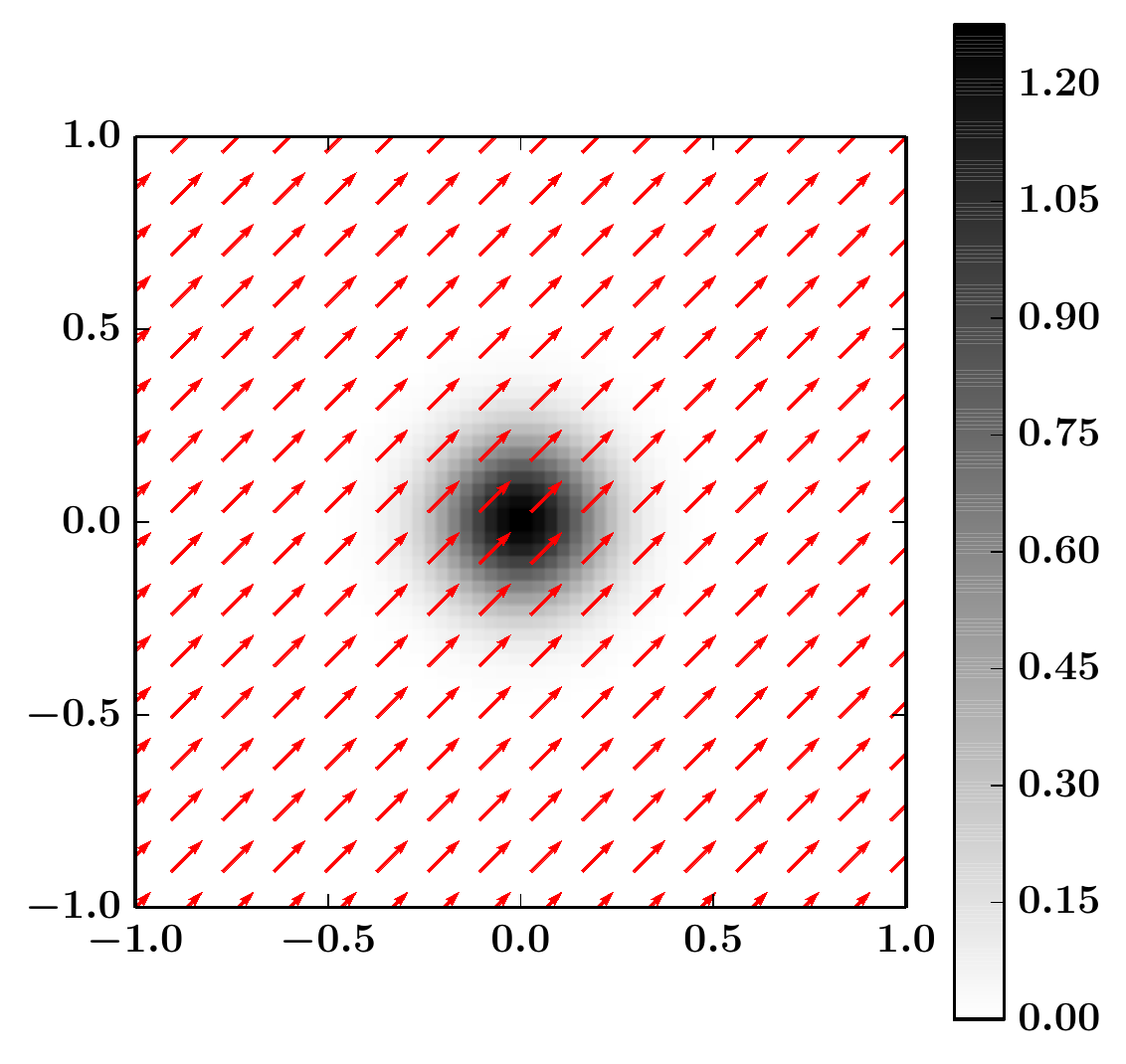}
  \includegraphics[width=8cm]{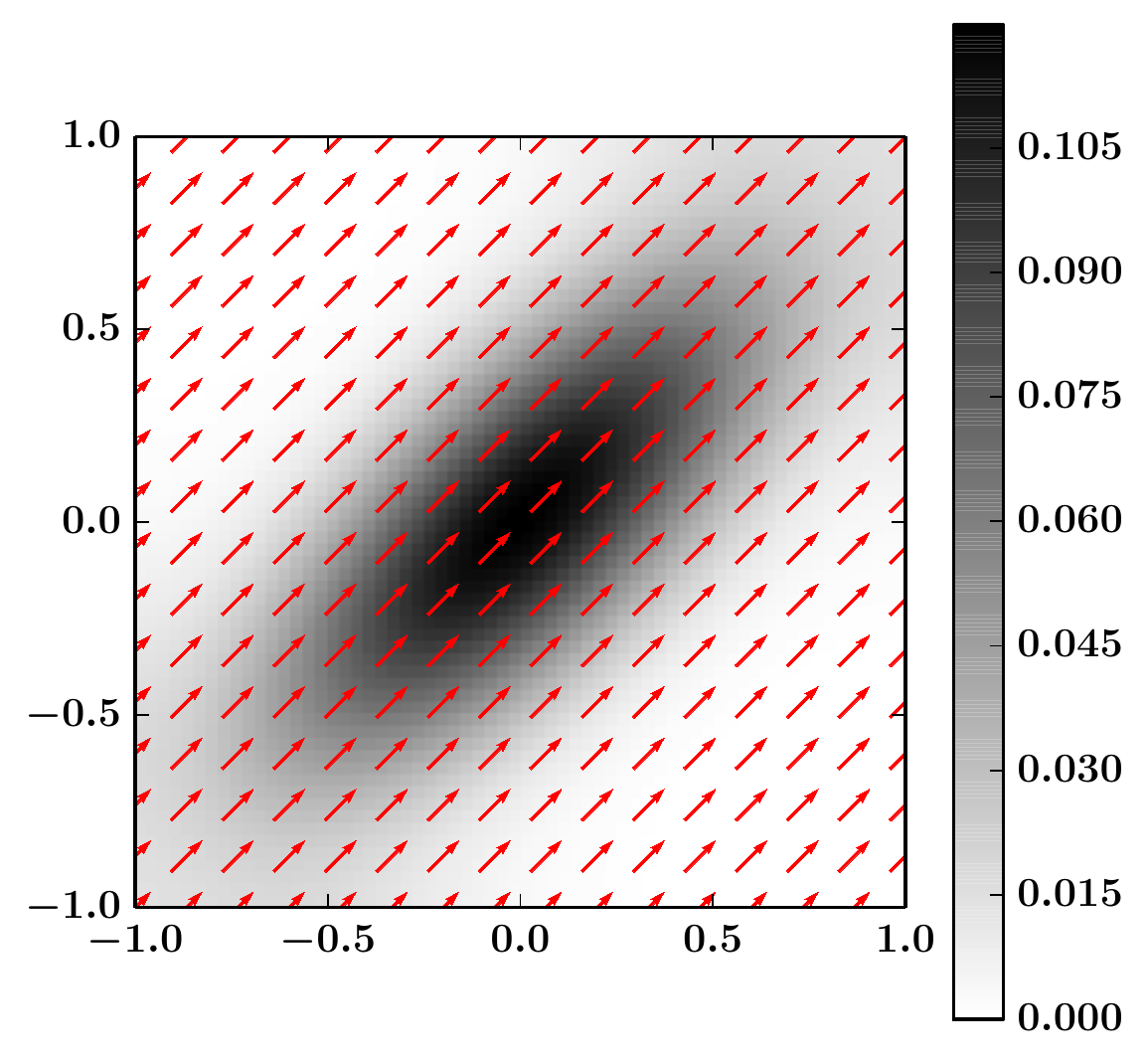}\\
  \includegraphics[width=8cm]{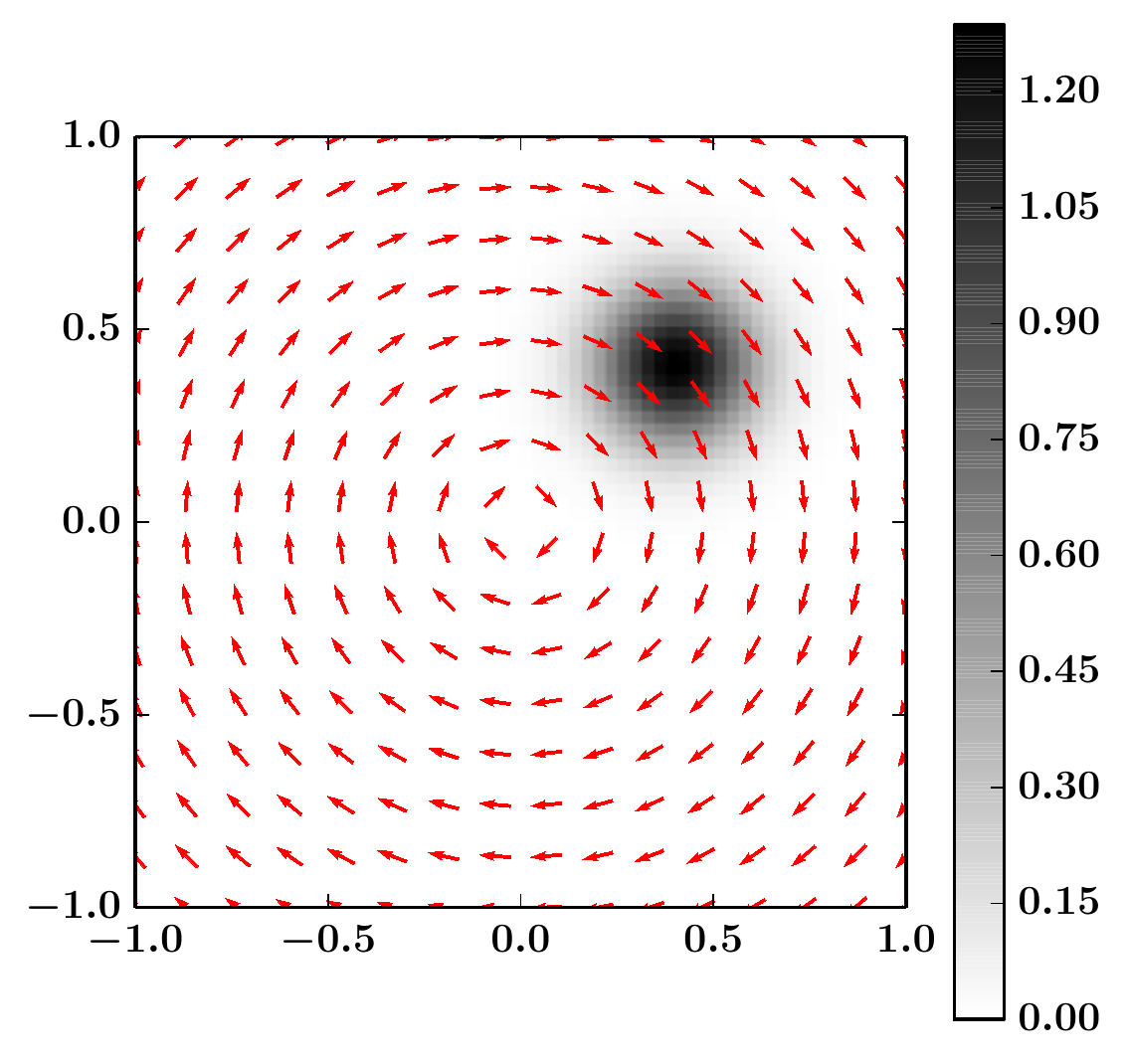}
  \includegraphics[width=8cm]{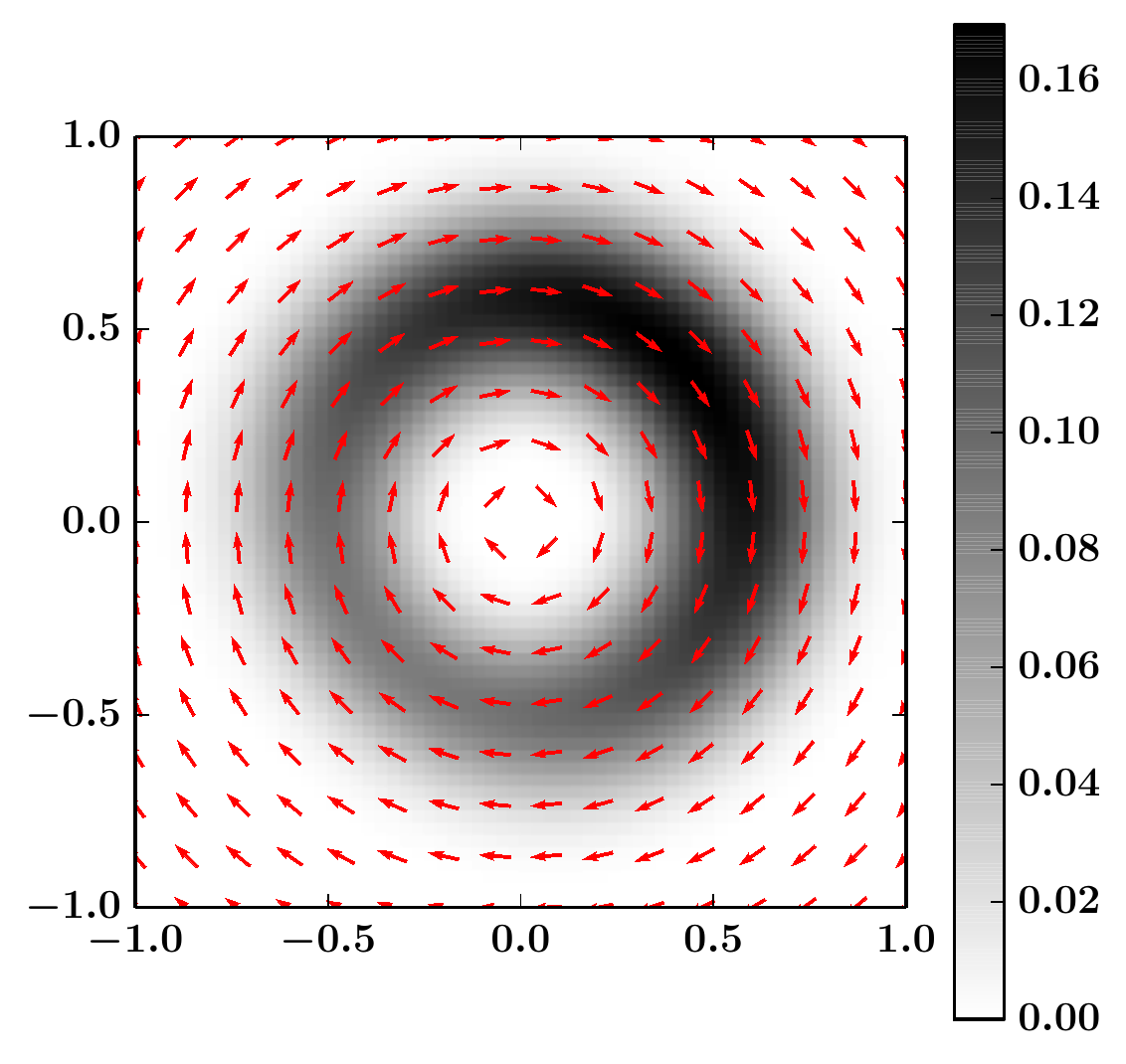}
  \caption{Diffusion tests for the new solver. The grey data field shows a slice of the total CR energy density, the arrows indicate the direction of the magnetic field. \textsc{Top}: Anisotropic diffusion along the magnetic field lines for a diagonal magnetic field configuration at time $0$ (left) and later (right). \textsc{Bottom}: Same as top row but for a circular magnetic field.}
  \label{fig:test-diffusion-aniso}
  \end{minipage}
\end{figure*}

As the diffusion is anisotropic with different diffusion coefficients for the component parallel and perpendicular to the magnetic field direction, we set up two different magnetic field configurations to test the diffusion in three dimensions. In the first setup (\ref{fig:test-diffusion-aniso}, top row) the magnetic field vectors are diagonal. The ratio of the parallel to the perpendicular diffusion coefficient is set to $K_\parallel/K_\perp = 10$. In the second setup (lower row of figure~\ref{fig:test-diffusion-aniso}) the magnetic field lines are circular $\vecB = (0,\hat{z},-\hat{y})$. 

\section{Adiabatic Losses}%
\label{sec:ad-losses-tests}

\subsection{Idealised spectrum}

Let us ignore the effects of diffusion for now and only consider the evolution of the spectrum under energy losses \citep[see, e.g.][]{Longair2011},
\begin{equation}
  \frac{dN(E)}{dt} = \frac{\partial}{\partial E}\ekl{b(E)N(E)} + Q(E).
\end{equation}
Here $b(E)$ describes the losses and $Q(E)$ the CR source term. Assuming an infinite uniform source with an injection spectrum of the form $Q(E) = \kappa E^{-p}$, the spatial dependencies disappear and the equation reduces to
\begin{equation}
  \label{eq:loss-equation}
  \frac{d}{dE}\ekl{b(E)N(E)} = -Q(E).
\end{equation}
Assuming $N(E)\rightarrow0$ as $E\rightarrow\infty$, we can integrate equation~(\ref{eq:loss-equation}) to give
\begin{equation}
  N(E) = \frac{\kappa E^{-(p-1)}}{(p-1)b(E)}.
\end{equation}
Adiabatic losses scale as $b(E)\propto E$, which results in an unchanged spectrum $N(E)\propto E^{-p}$.

\subsection{Numerical limits}
\begin{figure}
  \centering
  \includegraphics[width=8cm]{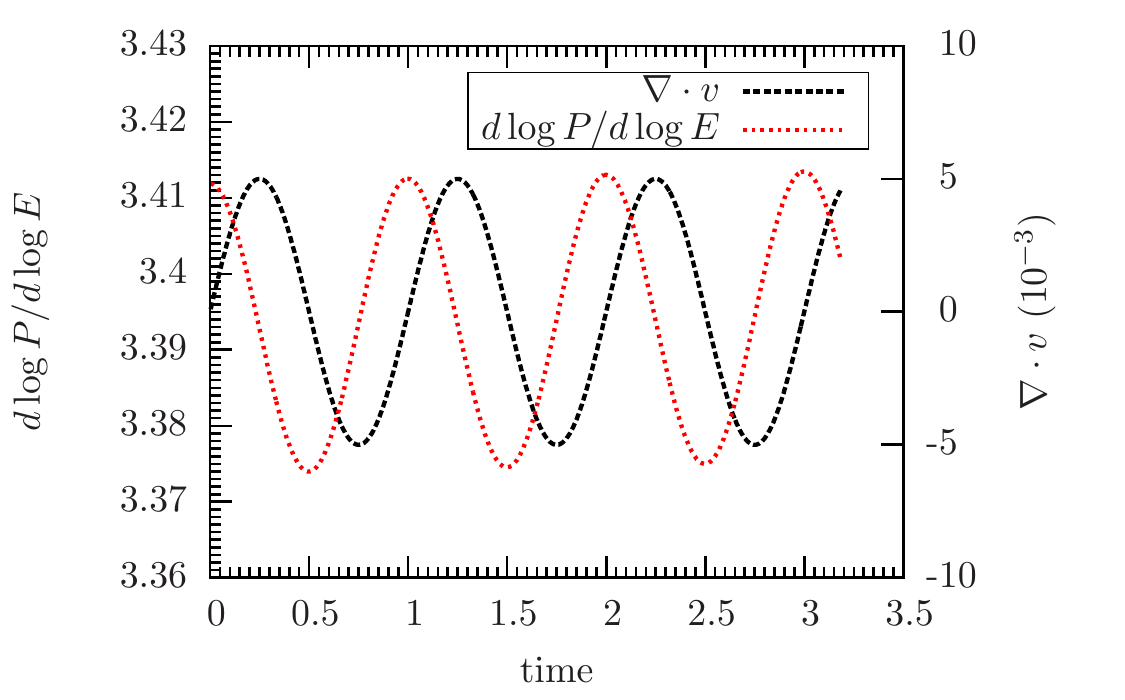}
  \includegraphics[width=8cm]{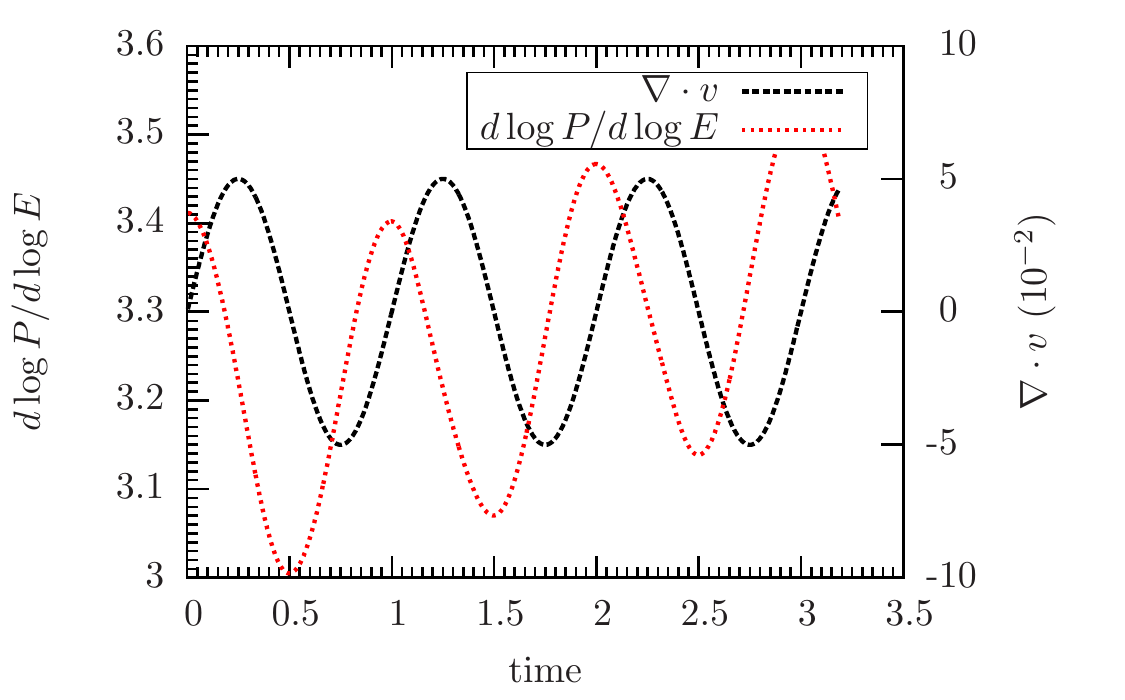}
  \caption{Periodic perturbation in $\nabla\cdot v$ and spectral response. For small perturbations the spectral changes are reversible (upper plot), for larger $\nabla\cdot v$ the reversibility is affected. The simulations have values in between the two cases shown above and are locally marginally affected. Globally, however, the adiabatic losses do not play a dominant role.}
  \label{fig:ad-losses-sine-perturbation}
\end{figure}

This simple, often cited statement of an unchanged spectrum for adiabatic losses does not hold in our simulations for two reasons. First, we do not have a continuous energy input but one distinctive point in time, at which the energy is injected. The second reason is given by the spectral limits of our code. We include ten fixed energy bins for the cosmic rays with a fixed minimum and maximum energy. In order to conserve energy in the simulations CRs do not enter nor leave the spectral energy boundaries. All shifts in the spectrum thus need to accumulate at the highest or lowest bin. A continuous compression for instance results in all energy pushed to the highest energy bin which does not give a proper spectrum any more. In order to verify that our implementation of the adiabatic losses works, we can not rely on an unchanged spectral index but need to show the conservation of the total energy and the reversability of the energy flow. Energy is conserved by construction of the fluxes between energy bins, so energy is conserved up to machine precision for each cell. The reversability of the adiabatic losses is slightly more complicated. One additional limitation is the piecewise constant distribution function. The energy fluxes between two bins are only exactly reversible if neighbouring values of the distribution function are equal $f_i=f_{i+1}$ or in the limit of infinitely many bins. A better result could be achieved if the distribution function is solved for with a more sophisticated form, e.g. with a piecewise power-law function. However, this means that equation~(\ref{eq:average-CR-energy}) depends on the values $f_i$, and the values for the distribution function need to be found iteratively, which is numerically much more expensive. We could also increase the energy range, such that the range of CR energy we put in does not reach the highest/lowest energy bin. However, including more bins strongly increases the computational cost. Increasing the range of each bin would keep the computational cost but would strongly coarsen the energy resolution of the dynamically interesting range. Figure~\ref{fig:ad-losses-sine-perturbation} shows the change of the slope under a sinusodial perturbation in $\nabla\cdot\mathbf{v}$ for a weak (upper plot) and strong (lower plot) amplitude. If the logarithmic spectral slope changes by roughly $0.5$ over one period of the perturbation, the spectral slopes are not reversible and show deviations caused by the asymmetry in the distribution function. However, we note that the sinusodial perturbation reverses the sign six times in our test, which is unlikely to happen in our simulation setups.

Running the same test with the physical numbers of the simulation such that the three periods correspond to the simulation time of $t=100\,\kyr$ shows the transition, at which the reversibility is lost, between $\nabla\cdot v \sim 10^{-12}\,\second^{-1}$ and $\nabla\cdot v \sim 10^{-11}\,\second^{-1}$. The simulation outside the SN shell region shows absolute values below $|\nabla\cdot v| < 10^{-13}\,\second^{-1}$, the shock regions reach absolute values ranging from $|\nabla\cdot v| \in 2-8 \times 10^{-12}\,\second^{-1}$. The SN shell is thus just at the limit, where the reversibility of the numerical methods shows deviations. However, as the region ahead of the shock is only exposed to the compression once, the reversibility problem does not apply for the regions far ahead of the shock, where the CRs transfer energy to the gas. The region behind the shock might be affected, but does not contribute to the enhancement of the kinetic energy ahead of the shock. In principle, the region behind the shock can harden the spectrum irreversibly, which allows the CRs to diffuse faster and reach the distances at which the CRs deposit the additional kinetic energy. However, from figure~\ref{fig:spectra-comp-ad-noad} we see that adiabatic losses -- whether in a $\nabla\cdot v$ regime that is reversible or not -- mainly enhance the high-energy CRs, which are not the main driver of the acceleration, cf. figure~\ref{fig:S1-B2-f1-new-spec-gradient-ratio}.

\subsection{SN explosion in one cell}%

\begin{figure}
  \centering
  \includegraphics[width=8cm]{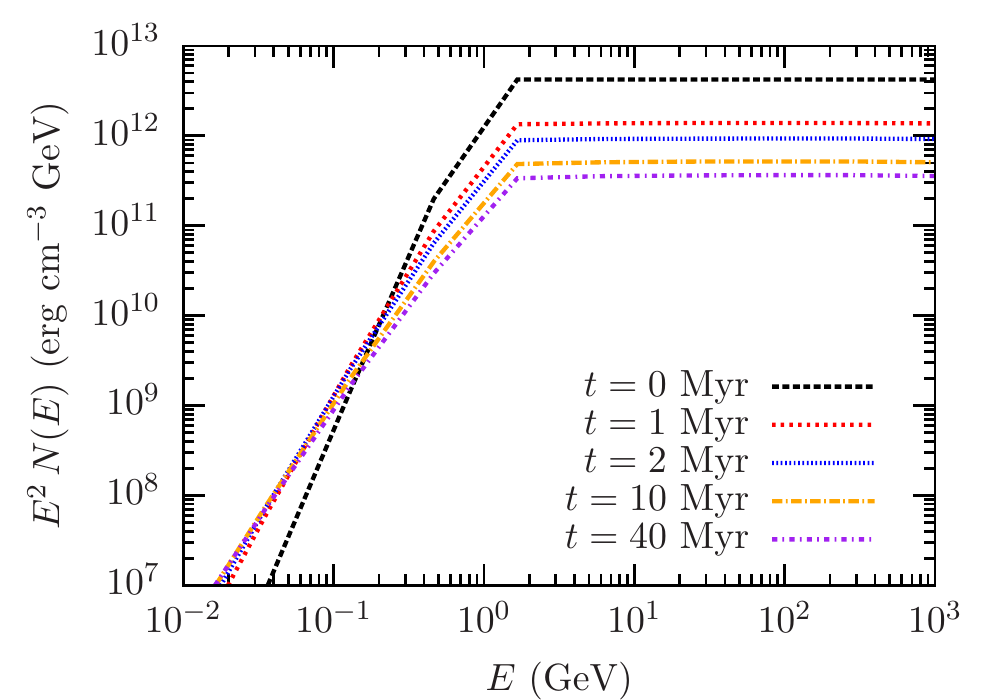}
  \caption{Time evolution of the CR spectrum for a test run with a SN injection region of only one cell. The low resolution does not allow for shock effects in the interior of the injection region and the spectrum is not hardened over time, cf. figure~\ref{fig:spectra-comp-ad-noad}.}
  \label{fig:tiny-SN-spectrum-time}
\end{figure}

By injecting CRs into a computational region with many cells, we resolve the shocks in the interior of the injection region and get reverse shocks that harden the spectrum in the centre of the SN injection region. In order to investigate the effects of the expansion on the spectrum we perform a simulation, where we inject the SN energy in only one cell at the centre of the domain. In order to avoid numerical instability by injecting $10^{51}\,\erg$ in one single cell, we reduce the total injected energy $10^{48}\,\erg$. The expansion of the shell will thus be slower and simulation time longer. We switch off the CR diffusion and measure the total energy and the spectrum in the center over time. Figure~\ref{fig:tiny-SN-spectrum-time} shows the spectrum at the centre of the box over time. The total energy drops and the spectral shape at the high-energy range stays unchanged during the expansion phase. The low-energy part of the spectrum is changed as the energy needs to accumulate there.

\end{appendix}

\clearpage
\newpage
\bibliographystyle{apj}
\bibliography{astro.bib}

\end{document}